\documentclass[useAMS,usenatbib]{mn2e}
\usepackage{footnote,graphicx,natbib,color,multirow,amsmath,url,amssymb,tabularx,amssymb, mathtools, listings, float}
\usepackage{aas_macros}

\def\lesssim{\mathrel{\hbox{\rlap{\hbox{\lower3pt\hbox{$\sim$}}}\hbox{\raise2pt\hbox{$<$}}}}}

\definecolor{check}{rgb}{0,0,0}

\definecolor{refer}{rgb}{0,0,0}
\def\referee		{\color{refer}}

\definecolor{refer2}{rgb}{0,0,0}
\def\refereeii		{\color{refer2}}

\begin{document}

\title[\textsc{snitch}: inferring informative  SFHs]{SNITCH: Seeking a simple, informative star formation history inference tool}
\author[Smethurst et al. 2019]{R. ~J. ~Smethurst,$^{1,2}$ M.~Merrifield,$^{1}$ C. ~J. ~Lintott,$^{2}$ K.~L.~Masters,$^{3}$ B.~D.~Simmons,$^{4,5}$ \newauthor A.~Fraser-McKelvie,$^{1}$ T.~Peterken,$^{1}$ M.~Boquien,$^{6}$ R.~A.~Riffel,$^{7}$ N.~Drory$^{8}$
\\ $^1$ School of Physics and Astronomy, The University of Nottingham, University Park, Nottingham, NG7 2RD, UK
\\ $^2$ Oxford Astrophysics, Department of Physics, University of Oxford, Denys Wilkinson Building, Keble Road, Oxford, OX1 3RH, UK
\\ $^3$ Institute of Cosmology and Gravitation, 
University of Portsmouth, Dennis Sciama Building, Barnaby Road, Portsmouth, PO13FX, UK
\\ $^4$ Physics Department, Lancaster University, Lancaster, LA1 4YB, UK
\\ $^5$ Center for Astrophysics and Space Sciences (CASS), Department of Physics, University of California, San Diego, CA 92093, USA
\\ $^6$ Universidad de Antofagasta, Centro de Astronom\'ia, Avenida Angamos 601, Antofagasta, 1270300, Chile
\\ $^7$ Departamento de Física, Centro de Ciências Naturais e Exatas,  Universidade Federal de Santa Maria, Av. Roraima 1000,\\ CEP 97105-900, Santa Maria, RS, Brazil 
\\ $^8$ McDonald Observatory, The University of Texas at Austin, 1 University Station, Austin, TX 78712, USA
\\
\\Accepted 2019 January 21. Received 2019 January 17; in original form 2018 October 12
}

\maketitle

\begin{abstract}
Deriving a simple, analytic galaxy {\refereeii star formation history} (SFH) using observational data is a complex task without the proper tool to hand. We therefore present \textsc{snitch}, an open source code written in \emph{Python}, developed to quickly ($\sim2$ minutes) infer the parameters describing an analytic SFH model from the emission and absorption features of a galaxy spectrum {\refereeii dominated by star formation gas ionisation}. \textsc{snitch} uses the Flexible Stellar Population Synthesis models of \cite{conroy09}, the MaNGA Data Analysis Pipeline and a Markov Chain Monte Carlo method in order to infer three parameters (time of quenching, rate of quenching and model metallicity) which best describe an exponentially declining quenching history. This code was written for use on the MaNGA spectral data cubes but is customisable by a user so that it can be used for any scenario where a galaxy spectrum has been obtained, and adapted to infer a user defined analytic SFH model for specific science cases. Herein we outline the rigorous testing applied to \textsc{snitch} and show that it is both accurate and precise at deriving the SFH of a galaxy spectra. The tests suggest that \textsc{snitch} is sensitive to the most recent epoch of star formation but can also trace the quenching of star formation even if the true decline does not occur at an exponential rate. With the use of both an analytical SFH and only five spectral features, we advocate that this code be used as a comparative tool across a large population of spectra, either for integral field unit data cubes or across a population of galaxy spectra. 
\end{abstract}


\section{Introduction}

{\referee Significant insight into the star formation history (SFH) of a galaxy can be obtained from its measured spectral features \citep{kauffmann03b, dressler04, li15, wang18, zick18}, such as specific emission lines and Lick absorption indices \citep{burstein84, faber85, burstein86, gorgas93, worthey94b, trager98}. Similarly, the parametrisation of a galaxy's complex SFH into a simple analytic form, has informed many of the mechanisms which drive the evolution of galaxies across cosmic time \citep{tinsley72, gavazzi02, martin07, kriek10, oemler13, schawinski14, simha14, abramson16, smethurst15}. Combining these two approaches by using spectral features to infer a parametrised SFH has recently allowed for further understanding of this complex problem \citep{nog18, zick18}.}

{\referee This method is complimentary to that of full spectral fitting which utilises all the information available from an observation and allows the determination of a comprehensive evolutionary history of a galaxy spectra including age, metallicity, mass-to-light ratio and SFH. Such a fit is often performed with an un-parametrised SFH so that the resulting SFH is composed of many sharp bursts of single stellar populations (SSPs) which are not always informative in specific science cases (for example, investigating starburst galaxies, post-starbursts or quenching galaxies). Whilst there are many publicly available codes which provide a full spectral fit to a galaxy spectrum \citep[SFH;][]{cappellari04, heavens04, cidfernandes05, ocvirk06, tojeiro07, noll09, conroy14, chevallard16, wilkinson17}, there are few that provide the targeted inference of a parametrised SFH for a user's specific science case given measured spectral features. Such a method does not return a comprehensive evolutionary history of a galaxy like a full spectral fit (since the majority of age and metallicity sensitive information is found in the continuum; $\sim75\%$ \citealt{chill09, chill11}), but it does allow for the derivation of comparative, informative SFHs across a population of galaxy spectra.}

{\referee This method is particularly attractive} with the recent influx of data from integral field unit (IFU) surveys targeting the internal dynamics and structure of large samples of galaxies, such as MaNGA \protect\citep[Mapping Nearby Galaxies at Apache Point Observatory;][]{bundy15}, SAMI \protect\citep[Sydney-AAO Multi-object Integral-field spectrograph;][]{bryant12} and CALIFA \protect\citep[Calar Alto Legacy Integral Field spectroscopy Area survey;][]{sanchez12}. Rather than obtaining a single spectra per galaxy, these surveys acquire multiple spectra per galaxy using {\referee configurations} of over 100 fibres. 

MaNGA \citep{bundy15} is an integral-field spectroscopic survey of 10,000 galaxies undertaken by the fourth phase of the Sloan Digital Sky Survey, SDSS-IV; \cite{blanton17}. The expectation is that over $100,000$ spectra will be obtained by MaNGA. Whilst this is not an unreasonable number of galaxy spectra (the Main Spectroscopic Galaxy Sample of the Sloan Digital Sky Survey totalled roughly $10^6$ spectra; \citealt{strauss02}) {\referee deriving comprehensive evolutionary histories for these will be time consuming and complex, a feat which some groups in MaNGA have already begun to undertake (see work on Pipe3D by \citealt{sanchez16} and on FIREFLY by \citealt{goddard17}). Although the products from these full spectral fitting routines are incredibly valuable and numerous, they are not always appropriate for all science cases.} 

We therefore present the open source \emph{Python} software package, \textsc{snitch}\footnote{\url{http://www.github.com/rjsmethurst/snitch/}}, which uses Bayesian statistics and a Markov Chain Monte Carlo (MCMC) method to quickly infer three parameters describing an analytical quenching SFH using a total of five absorption and emission spectral features which are sensitive to either star formation, age or metallicity. {\referee With the use of both an analytic SFH model and specific spectral features}, \textsc{snitch} is best suited to deriving the relative SFH parameters across a large sample of galaxy or IFU spectra in order to compare differences across the population. We do not recommend using \textsc{snitch} in order to quote the SFH parameters of only a single spectrum due to the generalising nature of an analytical SFH model {\referee and the loss of the age and metallicity sensitive information contained in the continuum.}  The benefits of using \textsc{snitch} include its adaptability to a particular targeted science case, a reduction in the time it takes to derive a specific analytic SFH for a large sample of galaxy spectra and the ease of comparing the resulting SFH parameters inferred for different spectra. 

This code has been developed originally for use with MaNGA integral field unit (IFU) spectral data cubes, however it can be used for any spectra where measurements of the absorption and emission (dominated by star formation gas ionisation) features are possible. Specifically \textsc{snitch} has been developed to study the quenching histories within spatially resolved regions of MaNGA galaxies, therefore herein we have defined a physically motivated SFH model parametrised by the time and rate that quenching occurs. However the SFH model used by \textsc{snitch} may be adapted by a user depending on the specific science case. For example, if a user wished to study starburst galaxies the SFH could be changed accordingly to parametrise the time and strength of the burst. 


Herein we describe \textsc{snitch} in Section~\ref{sec:code}, the expected output of the code in Section~\ref{sec:output}, along with the rigorous testing procedures applied to \textsc{snitch} in Section~\ref{sec:test}. Where necessary we adopt the Planck 2015 \citep{planck16} cosmological parameters with $(\Omega_m, \Omega_{\lambda}, h) = (0.31, 0.69, 0.68)$. 

\section{Description of Code}\label{sec:code}

\textsc{snitch} takes absorption and emission spectral features and their associated errors as inputs, assumes a quenching SFH model and convolves it with a stellar population synthesis (SPS) model to generate a synthetic spectrum. The predicted absorption and emission spectral features are then measured in this synthetic spectrum which are used to infer the best fit SFH model using Bayesian statistics and an MCMC method. 

We describe this process below, first defining our analytical SFH model (Section~\ref{sec:sfh}), how we convolve this with SPS models to produce synthetic spectra (Section~\ref{sec:fsps}), how these spectra are then measured to provide predicted model spectral features (Section~\ref{sec:dap}), which spectral features were chosen to be used as quenching indicators (Section~\ref{sec:choosespf}) and how these are used to infer the best fit SFH given the input parameters (Section~\ref{sec:emcee}). 

\subsection{Star Formation History Model}\label{sec:sfh}

The parametrised quenching SFH used by \textsc{snitch} was first described in \cite{smethurst15} for use in the \textsc{starpy} code\footnote{\textsc{starpy} is the precursor to \textsc{snitch}, performing a similar inference of a quenching SFH model using only an optical and near-ultraviolet colour. It has previously been used to study the quenching histories of AGN host galaxies \citep{smethurst16}, group galaxies \citep{smethurst17} and fast- \& slow-rotators \citep{smethurst18}. The code is publicly available here: \url{https://github.com/zooniverse/starpy}}. {\referee We summarise the description from \cite{smethurst15} here. The quenching SFH of a galaxy can be modelled as an exponentially declining star formation rate (SFR) across cosmic time as:
\begin{equation}\label{sfh}
SFR =
\begin{cases}
I_{\rm{sfr}}(t_q) & \text{if } t \leq t_q \\
I_{\rm{sfr}}(t_q) \times \exp{\left( \frac{-(t-t_{q})}{\tau}\right)} & \text{if } t > t_q 
\end{cases}
\end{equation}
where $t_{q}$ is the onset time of quenching and $\tau$ is the timescale over which the quenching occurs.  A smaller $\tau$ value corresponds to a rapid quench, whereas a larger $\tau$ value corresponds to a slower quench. We assume that all galaxies form at $t=0~\rm{Gyr}$. At the point of quenching, $t_{q}$, the SFH is defined to have an $I_{\rm{sfr}}(t_q)$ which lies on the relationship defined by \citet[][Equation 1]{peng10} for the $sSFR(m,t_q)$, for a galaxy with mass, $m = 10^{10.27} M_{\odot}$. Previous works by \citep{weiner06, martin07, noeske07,schawinski14, smethurst15} have shown that this analytic SFH appropriately characterise quenching galaxies. For galaxies which are still star forming, this model assumes a constant SFR. The SFR at any given redshift, z (or time of observation, $t_{obs}$), can now be generated for any set of SFH parameters.}



\begin{figure*}
\centering
\includegraphics[width=\textwidth]{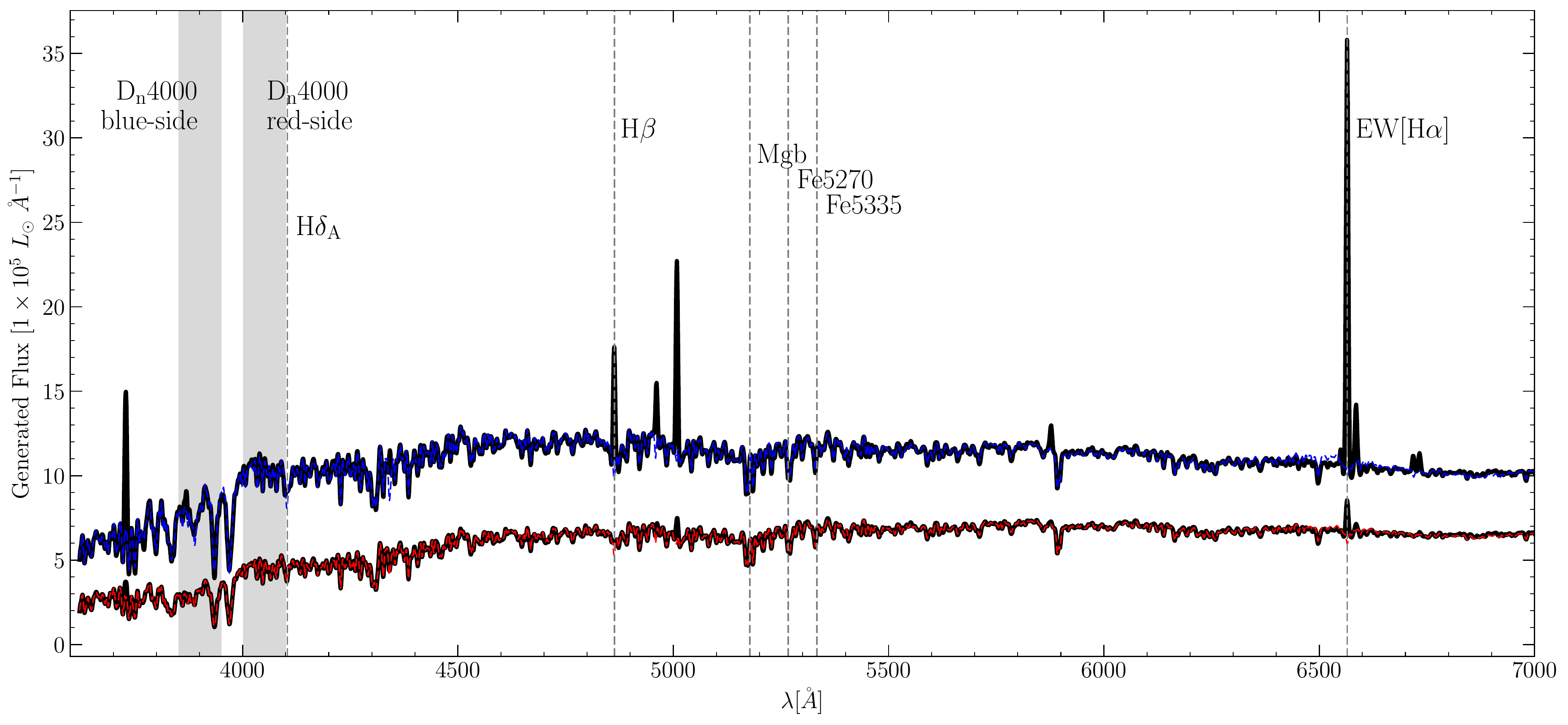}
\caption{Example synthetic spectra constructed using the Flexible Stellar Population Synthesis models of \citeauthor[][]{conroy09} (see Section~\ref{sec:fsps}), shown by the thick black solid lines, both with a SFH of $[Z,~t_q,~\tau]~=~[1~Z_{\odot},~10.0~\rm{Gyr},~0.5~\rm{Gyr}]$. Overlaid are the fits to the continuum returned by the MaNGA DAP (see Section~\ref{sec:dap}) shown by the blue dashed line for the spectra observed at $t_{\rm{obs}} = 10.4~\rm{Gyr}$, soon after quenching has begun, and the red dashed line for the spectra observed at $t_{\rm{obs}} = 13.8~\rm{Gyr}$, when the spectrum is quenched. \referee{We have also labelled each of the spectral features which are used as inputs for \textsc{snitch} (see Section~\ref{sec:choosespf}) and show their central wavelength by the dashed grey lines. The grey shaded regions show the blue- and red-side continuum regions used to measure the $\rm{D}_n4000$ feature.} See Section~\ref{sec:fsps}. }
\label{fig:spectrafit}
\end{figure*}


Whilst this is the SFH we have chosen to use, it is possible for a user to provide their own SFH function by adapting the \texttt{expsfh} function in \textsc{snitch}\footnote{\label{adapt}Information on how to adapt \textsc{snitch} for general usage is provided with the code in the GitHub repository: \url{http://www.github.com/rjsmethurst/snitch/}.}.

\subsection{Synthetic spectra generation}\label{sec:fsps}

We then employ SPS models in order to construct synthetic spectra for the SFHs defined in Section~\ref{sec:sfh}. These synthetic spectra will be measured in the same way as an observed spectrum (see Section~\ref{sec:dap}) in order to make a direct comparison using Bayesian statistics (see Section~\ref{sec:emcee}) to determine the `best fit' SFH model for a given spectral features input. 

In order to derive a realistic synthetic spectrum with our defined SFHs we used the Flexible Stellar Population Synthesis (FSPS)\footnote{\url{https://github.com/cconroy20/fsps}} code of \cite{conroy09} and \citet{conroy10}, which is written in \texttt{FORTRAN}, in conjunction with an existing \emph{Python} wrapper\footnote{\url{http://dfm.io/python-fsps/current/}} by \cite{python_fsps}. The FSPS \emph{Python} wrapper makes it possible to generate spectra (or magnitudes) for any arbitrary stellar population in just two lines of code.  

SPS methods rely on stellar evolution calculations to simulate all stages of stellar life, stellar spectral libraries, dust models and initial mass functions (IMFs) to translate the evolution of a hypothetical number of stars of varying ages and metallicities into a predicted integrated spectrum. FSPS also integrates \texttt{CLOUDY} \citep{ferland13} into its spectral output so that stellar emission lines can be synthesised along with the stellar continuum. {\referee Note that other sources of ionising radiation (e.g. {\refereeii from shocks or} active galactic nuclei; AGN) are not included in the spectral emission model in \textsc{snitch}. However, we note that if a user's specific science case has need for this, it would be possible to replace the spectral synthesis function (\texttt{generate\_spectra}) in \textsc{snitch} with one that does, {\referee ii in order to add another source of ionising photons}\footnote{See footnote~\ref{adapt}.}. With \textsc{snitch} in its standard form we encourage users to ensure that contaminating emission in an observed spectrum {\refereeii (from shocks or AGN)} has been removed or accounted for before using \textsc{snitch}. {\refereeii For example, either by fitting multiple components to spectral emission lines or by using a BPT diagram \citep{bpt} to exclude spectra dominated by AGN or shock ionisation. If these other sources of ionising radiation} are not accounted for, then the measured strength of emission due to star formation in an observed spectrum will be overestimated and the results obtained with \textsc{snitch} will be inaccurate.} {\refereeii This is of particular concern for users wishing to use \textsc{snitch} to study merging or interacting galaxies and starburst systems; \cite{rich14} showed that upward of $60\%$ of the total $\rm{H}\alpha$ emission flux in late-stage gas-rich merging ultraluminous and luminous infrared galaxies (U/LIRGs) is caused by radiative shocks. Similarly \cite{rich15} found that shocks account for up to $30\%$ of the total $\rm{H}\alpha$ emission flux in all interacting galaxies.}



In \textsc{snitch} we set up the FSPS models to produce spectra using the Padova isochrones \citep{girardi02} and MILES spectral library \citep{vazdekis16} with nebular emission, emission from dust \cite{draineli07}, a \cite{chabrier03} IMF and a \cite{calzetti00} dust attenuation curve. We also smooth the generated synthetic spectra to have the minimum velocity dispersion measurable by MaNGA, $77~\rm{km}~{s}^{-1}$ \citep{bundy15}. Spectra are generated for the 22 metallicities provided in the MILES models, ranging from $0.011~Z_{\odot}$ to $1.579~Z_{\odot}$ across a logarithmic age range spanning the Universe's history. FSPS does not allow for chemical enrichment of stellar birth material with time, i.e. the stellar populations have constant metallicity\footnote{Whilst we could attempt to provide a feature to implement chemical evolution modelling into these models this would firstly be full of uncertainty (the propagation of which would be unquantifiable {\referee unless one assumes a simplified case where no mergers are involved, e.g. see work by \citealt{kirby13, chill18}}) and secondly move us out of the regime of a simple, informative SFH model.}. {\referee The current version of FSPS does not allow for the $\alpha$-abundances of the stellar models to be varied. An investigation into how varying the $\alpha$-abundance would affect the generated synthetic spectra is out of the scope of this work}\footnote{{\referee We encourage the interested user to adapt the \texttt{generate\_spectra} function in \textsc{snitch} to take their own spectra generation code which does allow for the $\alpha$-abundance to vary from solar in order to investigate the impact on the measured spectral features. See footnote~\ref{adapt} for information on how to adapt \textsc{snitch} for general purpose.}}.

These spectra are generated across a logarithmically spaced 4-dimensional array in $[t_{obs},~Z,~t_q,~\tau]$ in order to facilitate faster run time during inference (see Section~\ref{sec:emcee}). These are generated for $15~t_{obs}$, $12~Z$, $50~t_q$, and $50~\tau$ values giving a grid of 450,000 synthetic spectra.

Two example synthetic spectra generated with FSPS for solar metallicity are shown by the solid black line in Figure~\ref{fig:spectrafit}. Note that FSPS generates spectra with flux units $L_{\odot}~\rm{Hz}^{-1}$, but that our spectral feature measurement procedure (see Section~\ref{sec:dap}) requires the flux in units of $\rm{\AA}^{-1}$. The spectra both have a SFH described by the parameters $[Z,~t_q,~\tau]~=~[1~Z_{\odot},~10.0~\rm{Gyr},~0.5~\rm{Gyr}]$. Overlaid are the fits to the continuum returned by the MaNGA DAP (see Section~\ref{sec:dap}) shown by the blue dashed line for the spectra observed at $t_{\rm{obs}} = 10.4~\rm{Gyr}$, soon after quenching has begun, and the red dashed line for the spectra observed at $t_{\rm{obs}} = 13.8~\rm{Gyr}$, when the spectrum is quenched.

 The fits are shown by the red dashed line for a spectra which has already quenched with $[Z,~t_q,~\tau] = [1~Z_{\odot},~11.5~\rm{Gyr},~0.1~\rm{Gyr}]$ and by the blue dashed line for a spectra which still has some residual star formation $[Z,~t_q,~\tau] = [1~Z_{\odot},~10.0~\rm{Gyr},~1.0~\rm{Gyr}]$ both observed at a redshift, $z=0.1$ (i.e. $t_{obs}=12.1~\rm{Gyr}$).

\subsection{Measuring the synthetic spectral features}\label{sec:dap}

This code was originally developed for a specific science case for use with MaNGA IFU data cubes. We therefore wished to measure our synthetic spectra generated using FSPS (see Section~\ref{sec:fsps}) in the same way as the MaNGA data. It is for that reason that we use the functions defined in the MaNGA Data Analysis Pipeline (DAP; Westfall et al. in prep. and Belfiore et al. in prep.) version $2.0.2$ in order to measure the features in our synthetic spectra. If the user has a predefined method for measuring emission and absorption features in their spectra, the \texttt{measure\_spec} function in \textsc{snitch} can simply be adapted\footnote{See footnote~\ref{adapt}.}.

Here we lay out the MaNGA DAP functions used in \textsc{snitch} to fit our synthetic spectra and obtain emission and absorption feature measurements for those unfamiliar:

\begin{enumerate}
\item pPXF \citep{cappellari04} is used to extract a fit to the stellar continuum of a full synthetic spectrum. Here we use the version of pPXF coded into the MaNGA DAP using the \texttt{PPXFFit} object and the MILES template spectral libraries. To do this we assume a `measurement' error on the synthetic spectra of $10\%$ of the generated flux value. 
\item Using the fit to the stellar continuum provided by pPXF, we measure the emission line features in a spectrum using the \texttt{Elric} object and the \texttt{"ELPFULL"} emission line database of all 26 lines provided in the MaNGA DAP. This procedure provides emission line fluxes, equivalent widths, and kinematics from single component Gaussian fits\footnote{{\referee The MaNGA DAP can provide both a Gaussian and non-parametric fit to the emission lines. Whilst the expectation is for the non-parametric fit to be more robust, analysis presented in upcoming work by Belfiore et al. (in prep) has shown that the Gaussian fit is appropriate for most spectra, except in the presence of broad line components (particularly of Type 1 AGN which make up only $1\%$ of the MaNGA sample).}}. All strong lines are fit, as well as the Balmer series up to $\rm{H}\epsilon$ and other weaker lines. 
\item We then measure the absorption indices in the emission line subtracted synthetic spectrum using the \texttt{SpectralIndices} object and the \texttt{"EXTINDX"} index database of all 42 indices provided in the MaNGA DAP. Spectral-index measurements including the $4000\rm{\AA}$ break, $\rm{TiO}$ bandhead features and the full Lick system. All indices are measured at the MaNGA resolution (specified for each index) and corrections are provided to a nominal, $\sigma_v = 0$ measurement. The measurements of the Lick indices are provided by convolving the MaNGA data to the Lick resolution. 
\end{enumerate}

When we use \textsc{snitch} we also apply the procedure outlined above to our observed spectra to obtain synthetic and measured spectral features with the same method. We encourage users to the same where possible, either by measuring their observed spectra using the MaNGA DAP functions coded into the \texttt{measure\_spec} function in \textsc{snitch} or by adapting this function to use a procedure defined by the user\footnote{See footnote~\ref{adapt}.}. {\referee It is imperative that the same spectral fitting procedure is applied to both the synthetic and observed spectra to negate the issue of model dependent emission line flux subtraction when measuring the absorption features. We note again that users should ensure that contaminating emission (e.g. {\refereeii from gas ionised by radiative shocks or AGN}) in their observed spectrum has been removed or accounted for before using \textsc{snitch} (see Section~\ref{sec:fsps}).}

\subsection{Choosing which spectral features to use}\label{sec:choosespf}

Whilst there are many star formation sensitive spectral features used previously in the literature \citep[see comprehensive review by][]{kennevans12} here we adopted a \emph{``first principles"} approach. We observed how each of the 26 emission and 42 absorption features measured by the MaNGA DAP (see Section~\ref{sec:dap}), changed across the model parameter space $[Z, t_q, \log \tau]$ with time of observation to determine which spectral features were most sensitive to SFR, metallicity and time of observation. 

\begin{figure}
\centering
\includegraphics[height=0.83\textheight]{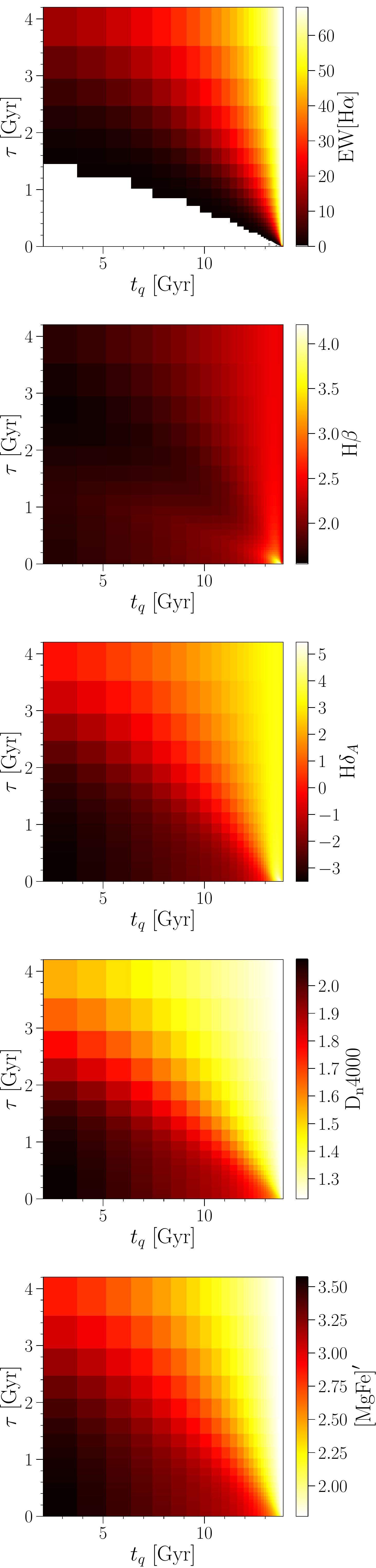}
\caption{The variation of model spectral features across the logarithmically binned two dimensional $[t_q, \log \tau]$ parameter space measured at $t_{obs}=13.8~\rm{Gyr}$ and solar metallicity, $Z=Z_{\odot}$. The features shown from top to bottom are the equivalent width of the $\rm{H}\alpha$ emission line and the spectral absorption indices $\rm{H}\beta$, $\rm{H}\delta_A$, $\rm{D}_n4000$ and $[\rm{MgFe}]^{\prime}$. Note that when a model has minimal star formation, the fitting code cannot measure an equivalent width of $\rm{H}\alpha$ therefore these values are masked out in the bottom left corner of the top panel. This figure shows how each feature is sensitive to the changing SFH and how they can be used to break the degeneracies that plague photometric studies of SFH. See Section~\ref{sec:choosespf}.}
\label{fig:rainbow}
\end{figure}

We looked at how plots similar to those shown in Figure~\ref{fig:rainbow} for all 26 emission features and 42 absorption features changed for at different ages and metallicities. This was not a blind selection, as parameters were labelled during this study, but all features were considered across the SFH model parameter space. Many features were degenerate with other stronger spectral features or did not show strong enough variation with a change in metallicity, age or quenching parameters, ruling them out as useful features for inference. We therefore selected the following features with which to infer the SFH parameters:
\begin{enumerate}[(i)]
\item The equivalent width (EW) of the $\rm{H}\alpha$ emission line, $\rm{EW}[\rm{H}\alpha]$, as it is the most sensitive to {\referee changes} in the current SFR;


\item $\rm{H}\beta$ absorption index, as it is most sensitive to any recent, rapid quenching that has occurred;

\item $\rm{H}\delta_A$ absorption index, as it is the most sensitive to A-stars and therefore star formation that has been cut off within the last $\sim1~\rm{Gyr}$;

\item $\rm{D}_n4000$ as it is most sensitive to older stars and therefore the age of the stellar population, however there is also an age-metallicity degeneracy for this feature so we also employ;

\item $\rm{[MgFe]}^{\prime}$ as it is most sensitive to the metallicity of the stellar population.

\end{enumerate}

{\referee Combining the spectral features listed above allows for all the different star formation timescales to be probed by using indicators of stellar populations of different ages. Of all the features listed above only $\rm{EW}[\rm{H}\alpha]$ and $\rm{D}_n4000$ are sensitive to the presence of dust (see \citealt{balogh99})} {\referee Unlike the $\rm{H}\alpha$ flux, which is sensitive to the current SFR, the EW[$\rm{H}\alpha$] measures the relative contribution of the $\rm{H}\alpha$ emission to the underlying continuum. Since the continuum is a proxy for stellar mass and the $\rm{H}\alpha$ emission arises around short-lived O and B stars, the EW[$\rm{H}\alpha$] is ideal for probing recent changes to the SFR in the past couple of $100$ Myr or so, in relation to the total integrated star formation over the galaxy’s lifetime \citep[see also][]{li15, zick18}.} It is worth noting that although these features were selected using this \emph{``first principles"} approach, they unsurprisingly appear frequently in many works studying galaxy SFRs and histories, e.g. \cite{kauffmann03, brinchmann04, goto05b, moustakas06, martin07, huang13, li15, wang18, spindler18, zick18} to name but a few. 

The variation in these five spectral features across the two dimensional $[t_q, \log \tau]$ SFH parameter space measured at $t_{obs}=13.8~\rm{Gyr}$ and solar metallicity, $Z=Z_{\odot}$ is shown in Figure~\ref{fig:rainbow}.

{\refereeii We note again here that only measurements from spectra dominated by emission due to gas ionisation from star formation should be input into \textsc{snitch} (see Section ~\ref{sec:fsps}). If a user has a spectra which they think may be contaminated, we recommend modelling for or removing this contamination before measuring the emission equivalent widths. If this is not possible, e.g. due to spectral resolution constraints, then we recommend omitting the $\rm{EW}[\rm{H}\alpha]$ measurement from the list of inputs to \textsc{snitch} (see below and Section~\ref{sec:missingtest}).}

Fewer than five spectral features can be provided to \textsc{snitch}, although not providing one of the five does restrict the accuracy to which a SFH can be inferred (see Section~\ref{sec:missingtest}). An estimate of the error on these measured values is also needed for \textsc{snitch} to run. The more precise the measurement of the spectral feature, the more precise the inferred SFH. It is possible for a user to adapt \textsc{snitch} to take any number of different spectral features which are appropriate for their scientific purpose\footnote{See footnote~\ref{adapt}}.

\subsection{Bayesian inference of SFH parameters}\label{sec:emcee}

For the SFH problem at hand, using a Bayesian approach requires consideration of all possible combinations of the model parameters $\theta \equiv [Z, t_{q}, \log \tau]$ (the hypothesis in this instance). Assuming that all galaxies formed at $t=0~\rm{Gyr}$, we can assume that the `age' of a spectrum is equivalent to an observed time, $t_{obs}$. We used this  `age' to calculate the five \emph{predicted} spectral features, $s,p$, at this cosmic time for a given combination of $\theta$, $\vec{d}_{s,p}(\theta, t_{obs}) = {s_p(\theta, t_{obs})}$. The predicted spectral features can now directly be compared with the five input \emph{observed} spectral features $\vec{d}_{s, o} = \{s_o\}$ which have an associated measurement error $\vec{\sigma}_{s, o} = \{\sigma_{s, o}\}$. For a single spectrum, the likelihood of a given model $P(\vec{d}_{s, o}|\theta, t_{obs})$ can be written as:


\begin{multline}\label{like}
P(\vec{d}_{s, o}|\theta, t_{obs}) = \prod_{s=1}^{S} P(s_{o}|\theta, t_{obs}) = \\ \prod_{s=1}^{S} \frac{1}{\sqrt{2\pi\sigma_{s,o}^2}} \exp{\left[ - \frac{(s_{o} - s_{p}(\theta, t_{obs}))^2}{\sigma_{s, o}^2} \right]},
\end{multline}

where $S$ is the total number of spectral features used in the inference. Here we have assumed that $P(s_{o}|\theta, t_{obs})$ are all independent of each other and that the errors on the observed features, $\sigma_{s, o}$, are also independent and Gaussian (a simplifying assumption but difficult to otherwise constrain). To obtain the probability of a set of $\theta$ values, i.e. a SFH model, given the observed spectral features: $P(\theta|\vec{d}_{s,o}, t_{obs})$, we use Bayes' theorem:
 \begin{equation}\label{eq:bayes}
P(\theta|\vec{d}_{s,o}, t_{obs}) = \frac{P(\vec{d}_{s,o}|\theta, t_{obs})P(\theta)}{\int P(\vec{d}_{s,o} |\theta, t_{obs})P(\theta) d\theta}.
\end{equation}
We assume the following prior on the model parameters so that the probability drops off at the edges of the parameter space: ${P(\theta)~=~1}$ if 0 $< Z [Z_{\odot}] \leq 1.5 \text{ and } 0 < t_q ~\rm{[Gyr]}~ \leq 13.8 ~ \text{ and } ~ 0 < \tau  ~\rm{[Gyr]}~ \leq 5.9$ and ${P(\theta)~=~2\times\exp\left(\log_{10}[5.9]\right)~-~\exp\left(\log_{10}[\tau]\right)}$ otherwise.


{\referee As the denominator of Equation~\ref{eq:bayes} is a normalisation factor, comparison between posterior probabilities for two different SFH models (i.e., two different combinations of $\theta = [Z, t_q, \log \tau]$) is equivalent to a comparison of the numerators of Equation~\ref{eq:bayes}. Markov Chain Monte Carlo (MCMC; \citealt{mackay03, emcee13, GW10}) analysis provides us with a robust method to compare the posterior probabilities for different $\theta$ values.

MCMC allows for an efficient exploration of the parameter space by avoiding areas with low likelihood. A large number of `walkers' are started at an initial position (i.e. an initial guess at the SFH, $\theta$) from which they each individually `jump' a randomised distance to a new position. If the probability in this new position is greater than the calculated probability at the original position then a `walker' accepts this change. Any new position then influences the direction of the  `jumps' of other walkers (true for ensemble MCMC only). This is repeated for a specified number of jumps after an initial `burn-in' phase. The length of this burn-in phase is determined after sufficient experimentation to ensure that the `walkers' have converged on the global minimum within the defined number of steps. Here we use \emph{emcee},\footnote{\url{dan.iel.fm/emcee/}} \citep{emcee13}, an affine invariant ensemble sampler written in \emph{Python} to explore the SFH parameter space for a given set of measured spectral features. \emph{emcee} returns the positions of the `walkers' in the predefined parameter space, which are analogous to the regions of high posterior probability, $P(\theta|\vec{d}_{s,o}, t_{obs})$.

For each run of \textsc{snitch}, the inference run is initialised with 100 walkers with a burn-in phase of 1000 steps before a main run of 200 steps. Acceptance fractions for each walker are difficult to estimate due to the fact that walkers often get stuck in local minima during a run (see Section~\ref{sec:pruning} for more information). 

}

\begin{figure}
\centering
\includegraphics[width=0.495\textwidth]{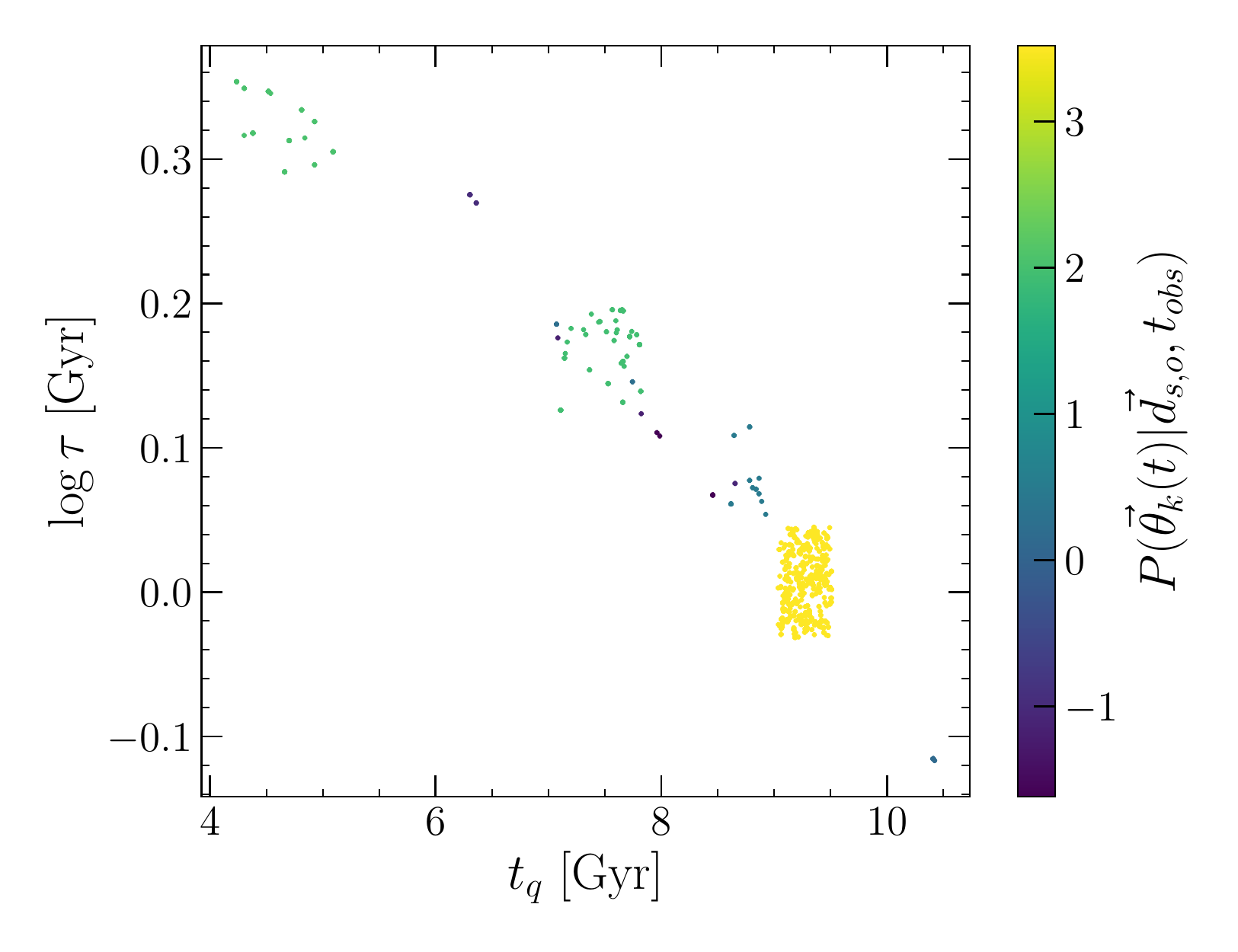}
\caption{This figure shows the walker positions marginalized over the $Z$ dimension into the two dimensional $[t_q, \log\tau]$ space and coloured by their characteristic $P(\vec{\theta}_k (t)|\vec{d}_{s,o}, t_{obs})$ value (see Equation~\ref{eq:bayes}). The higher the value of their log probability, the more likely the model is. The lower values of log probability for some groups of walkers suggests that these are indeed stuck in local minima. These clusters of walkers in local minima can be `pruned' (see Section~\ref{sec:pruning}) away to leave only the global minimum in the final output. Note that since we employ a nearest neighbours interpolation method across the look-up table (see Section~\ref{sec:emcee}) the resulting global minimum in parameter space traces the grid structure of the look-up table. See Section~\ref{sec:pruning}.}
\label{fig:localminima}
\end{figure}

\begin{figure*}
\centering
\includegraphics[width=0.495\textwidth]{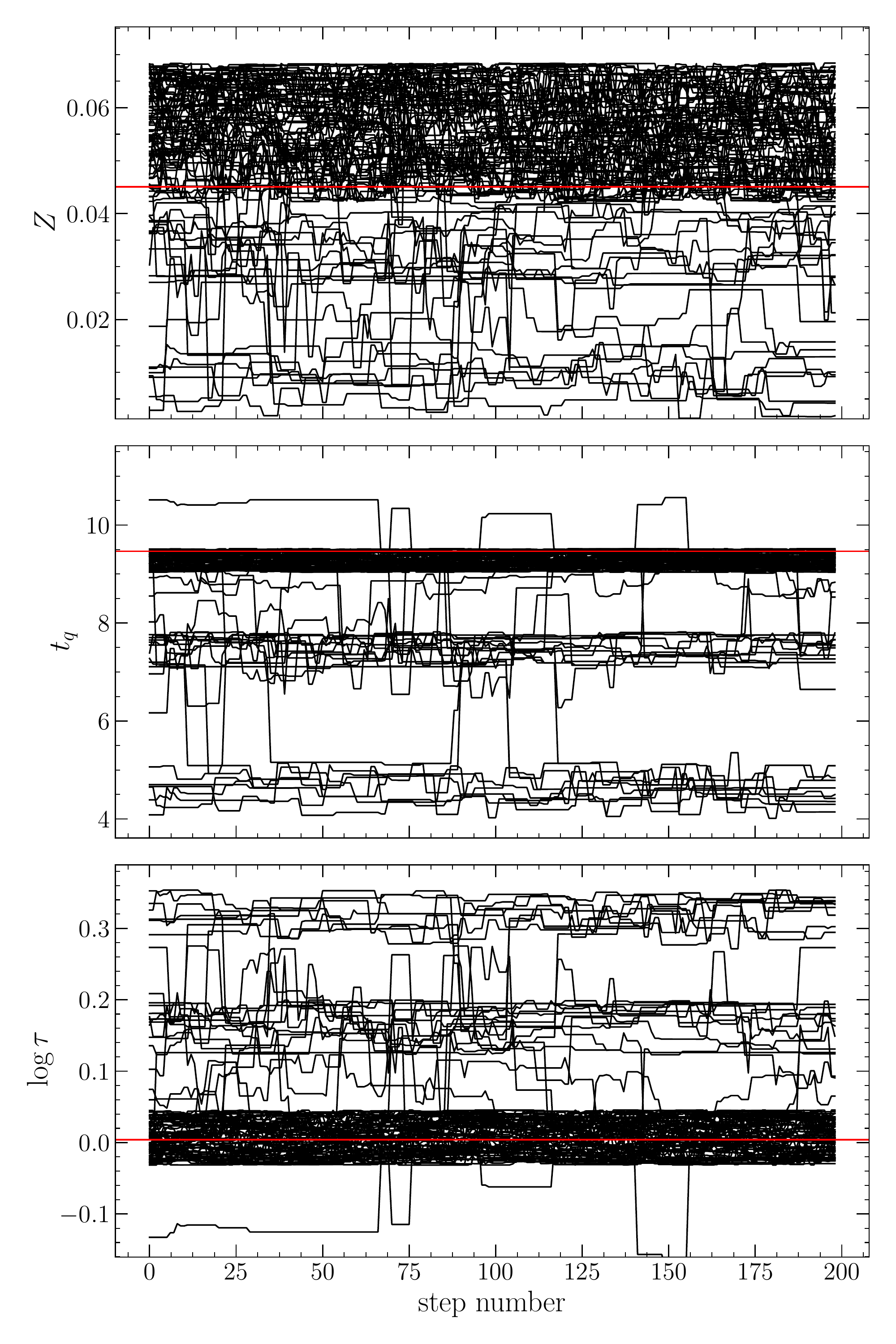}
\includegraphics[width=0.495\textwidth]{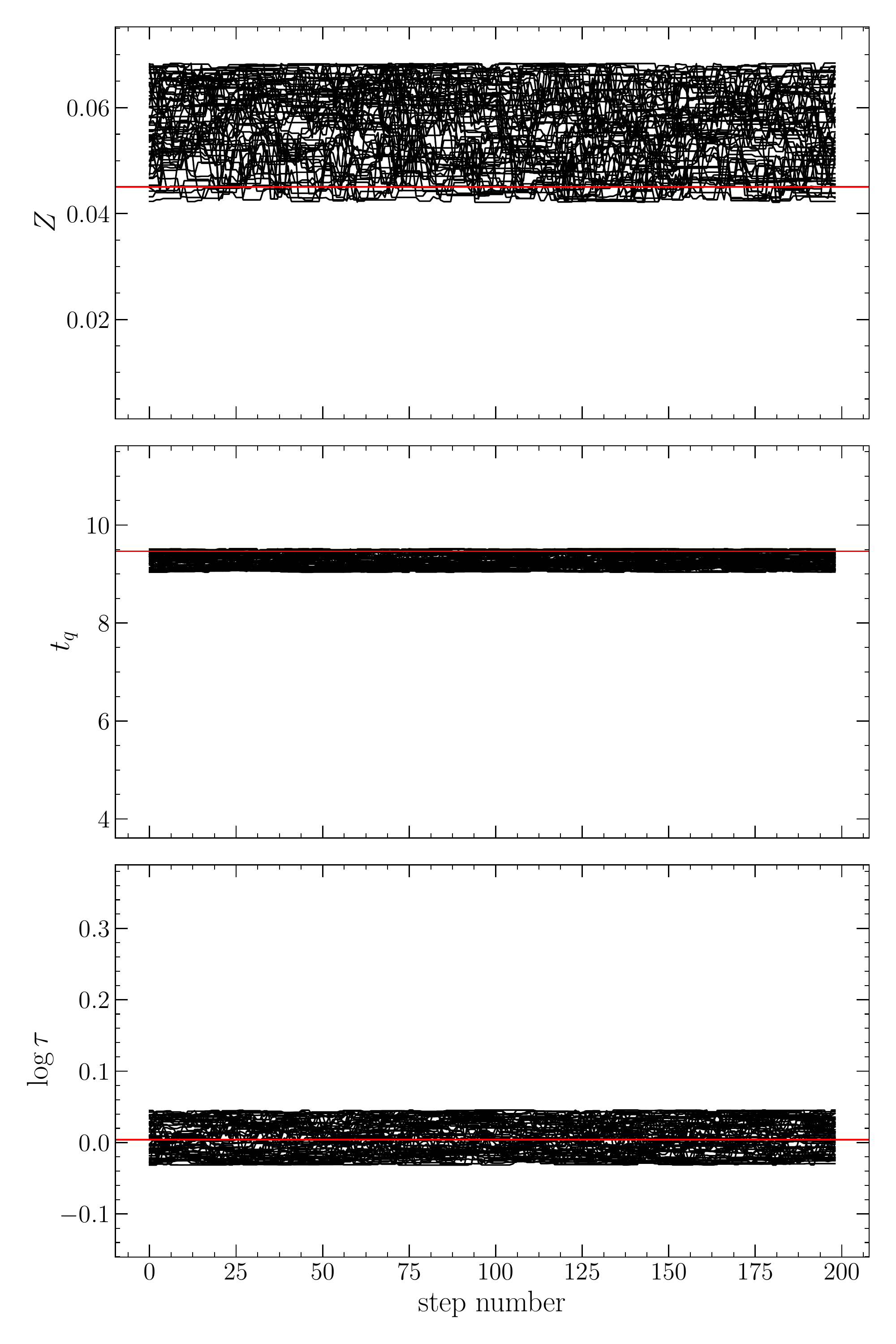}
\caption{The positions traced by the \emph{emcee} walkers with step number (i.e. time) in each of the $[Z, t_q, \log \tau]$ dimensions in the post burn-in phase before pruning (left) and after pruning (right). Walkers have got stuck in local minima (see also Figure~\ref{fig:localminima}) but some have managed to find the global minimum. The right panel therefore shows how the walkers left after pruning have fully explored the global minimum around the known true values (shown by the red lines in each panel). See Section~\ref{sec:output}.}
\label{fig:comparepruning}
\end{figure*}

With each `walker' jump to a new place in parameter space, a synthetic spectra must be generated then measured, as described in Section~\ref{sec:dap}, to produce predicted spectral parameters. Since this is very computationally expensive, a 4-dimensional look-up table of each of the five spectral parameters listed in Section~\ref{sec:choosespf} was generated across a logarithmically spaced grid in $[t_{obs},~Z,~t_q,~\tau]$\footnote{This look-up table will also be made publicly available for those users who want to use \textsc{snitch} in its original format. This is available in the GitHub repository \url{http://www.github.com/rjsmethurst/snitch/}}. We initialised our look-up table over a non-regular grid in order to optimise the number of useful $t_q$ values for each $t_{obs}$ value, i.e. quenched SFHs with $t_q \leq t_{obs}$. Those SFHs with $t_q > t_{obs}$ had constant SFR and so returned the same values for the spectral parameters regardless of the $t_q, \log\tau$ values. This allowed us to construct a finer array in $t_q$ for each value of $t_{obs}$ to pinpoint recent changes in the SFH more precisely. Figure~\ref{fig:rainbow} shows a slice in two dimensions of this look-up table, for $t_{obs} = 13.8~\rm{Gyr}$ and solar metallicity, $Z=Z_{\odot}$ for each of the five spectral parameters.

The look-up table is interpolated over (using a nearest neighbour approach to speed up run time over the irregular grid) to find spectral parameters for each `walker' jump to any new position in $[t_{obs},~Z,~t_q,~\log\tau]$ parameter space.

\subsection{Pruning walkers stuck in local minima}\label{sec:pruning}

After running \textsc{snitch} and inspecting the walker positions it became apparent that the walkers of \emph{emcee} would often get stuck in local minima. We therefore implemented a pruning method, as described in \cite{hou12}, in order to remove those walkers in local minima leaving only the global minima from which to derive inferred SFH parameters. The method outlined in \cite{hou12} is a simple one dimensional clustering method wherein the average negative log-likelihood for each walker is collected. This results in $L$ numbers; $\overline{l}_k$:
\begin{equation}\label{eq:lnumbers}
\overline{l}_k = \frac{1}{T} \sum^{T}_{t=1} P(\vec{\theta}_k(t)|\vec{d}_{s,o}, t_{obs}),
\end{equation}
where T is the total number of steps each walker, $k$, takes. $\vec{\theta}_k(t)$ is therefore the set of walker positions at a given step, $t$ in the MCMC chain. These $L$ numbers, $\overline{l}_k$, are therefore characteristic of the well which walker $k$ is in, so that walkers in the same well will have similar $\overline{l}_k$ (see Figure~\ref{fig:localminima} in which walkers are coloured by their characteristic $P(\vec{\theta}_k(t)|\vec{d}_{s,o}, t_{obs})$ value).

The walkers are all then ranked in order of decreasing average log likelihood, $\overline{l}_{(k)}$, or increasing $- \log \overline{l}_{(k)}$. If there are big jumps in the $- \log \overline{l}_{(k)}$, these are easy to spot and are indicative of areas where walkers have got stuck in local minima. The difference in $- \log \overline{l}_{(k)}$ for every adjacent pair of walkers is calculated. The first pair whose difference is a certain amount larger than the average difference previously is then identified like so:
\begin{equation}\label{eq:idprunes}
-\log \overline{l}_{(j+1)} + \log \overline{l}_{(j)} > Const − \frac{\log \overline{l}_{(j)} + log \overline{l}_{(1)}}{j - 1}.
\end{equation}

After some trial and error we decided on a constant value of $Const = 10000$. All the walkers with with $k>j$ are thrown away and only the ones with $k \leq j$ are kept after being identified as part of the global minimum. This can be seen in Figure~\ref{fig:comparepruning} wherein the walker positions at each step before pruning are shown in comparison to those after pruning in the main run stage. In the cases where the `walkers' did not get stuck in local minima, this pruning routine leaves the walker chains untouched.

\begin{figure*}
\centering
\includegraphics[width=0.8\textwidth]{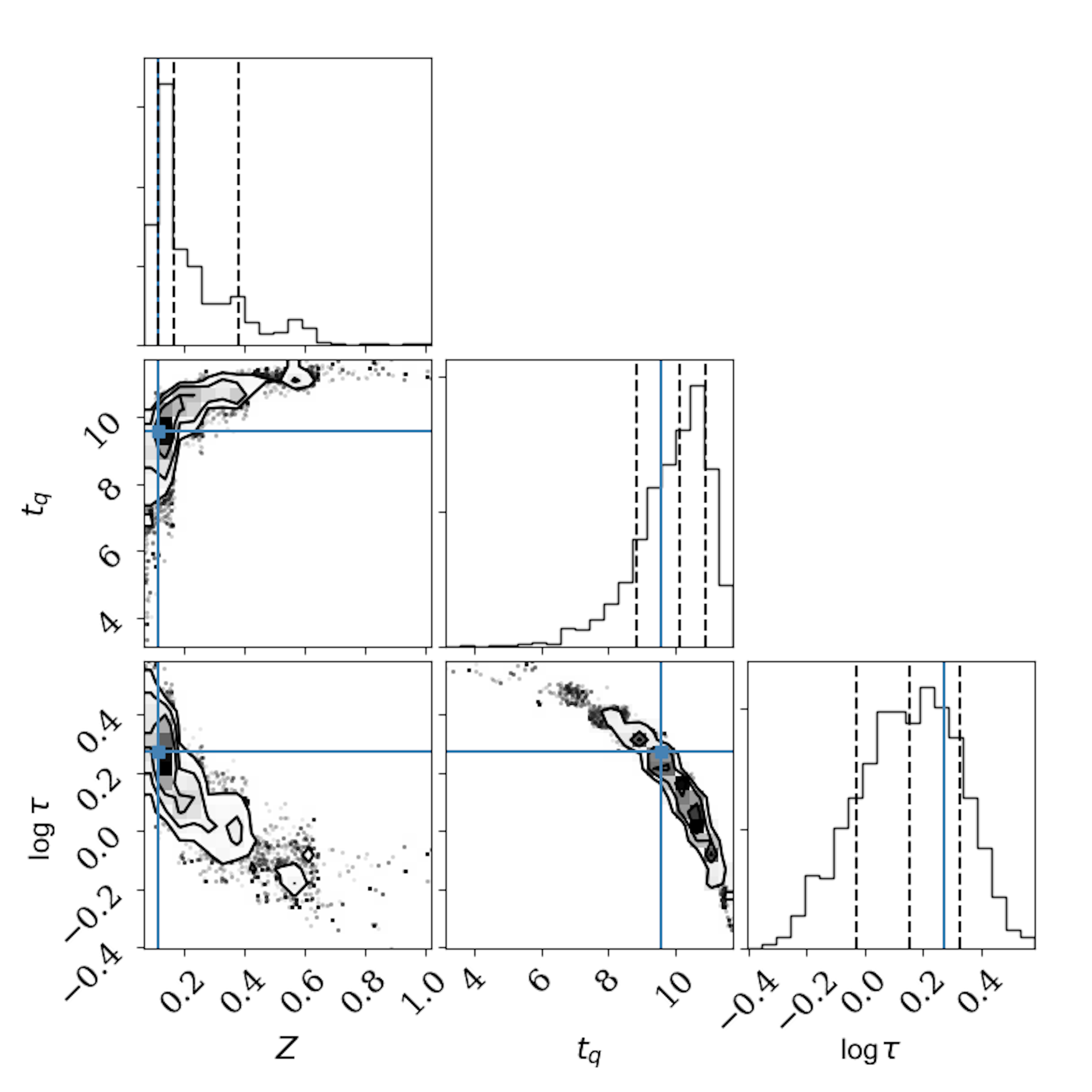}
\caption{Example output from \textsc{snitch} showing the posterior probability function traced by the MCMC walkers across the three dimensional parameter space $[Z, t_q, \log\tau]$. Dashed lines show the 18th, 50th and 64th percentile of each distribution function which can be interpreted as the `best fit' with $±1\sigma$. The blue lines show the known true values which \textsc{snitch} has managed to recover. See Section~\ref{sec:output}.}
\label{fig:output}
\end{figure*}

\section{Output of Code}\label{sec:output}


The burn-in and main run walker positions and posterior probabilities at each step are saved by \textsc{snitch}. From this three dimensional MCMC chain charting the $[Z, t_q, \log \tau]$ positions of the walkers around parameter space, the `best fit' $[Z, t_q, \log \tau]$ values along with their uncertainties can be determined from the 16th, 50th and 84th percentile values of the walker positions. These values are quoted by \textsc{snitch} at the end of a run. An example output from \textsc{snitch} for the predicted spectral features of a single known model SFH with a synthetic spectrum constructed with the FSPS models (see Section~\ref{sec:fsps}) is shown in Figure~\ref{fig:output}. This figure is also saved by \textsc{snitch} upon completion of a run.

The required inputs for \textsc{snitch} to run on a single spectrum are at least one, if not all, of $\rm{EW}[\rm{H}\alpha]$, $\rm{D}_n4000$, $\rm{H}\beta$, $\rm{H}\delta_A$ and $\rm{[MgFe]}^{\prime}$ and their associated errors and the galaxy redshift, $z$. To run \textsc{snitch} on a typical laptop on the spectral features of a single spectrum takes approximately 2 minutes. 

\section{Testing}\label{sec:test}

\begin{figure*}
\centering
\includegraphics[width=0.24\textwidth]{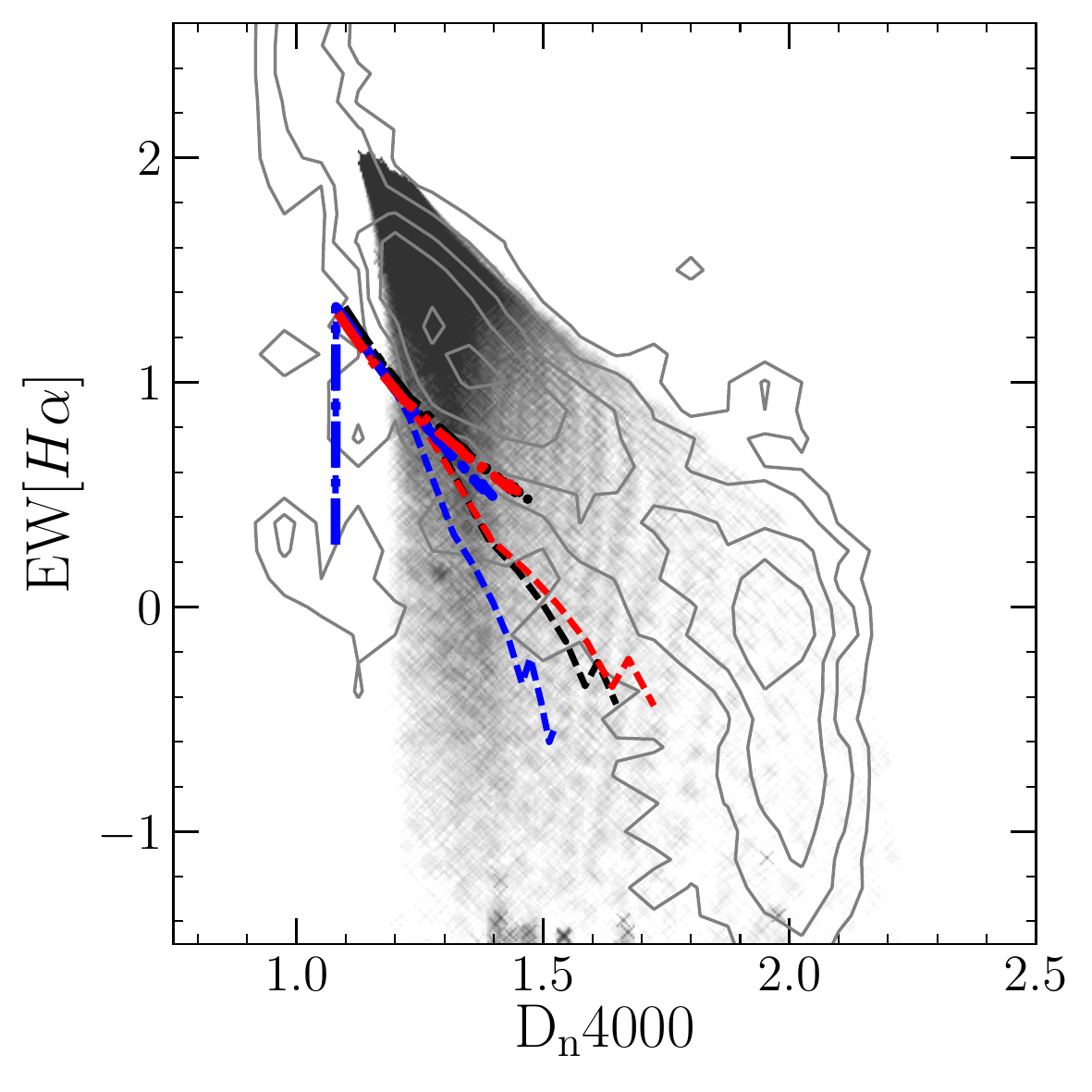}
\includegraphics[width=0.24\textwidth]{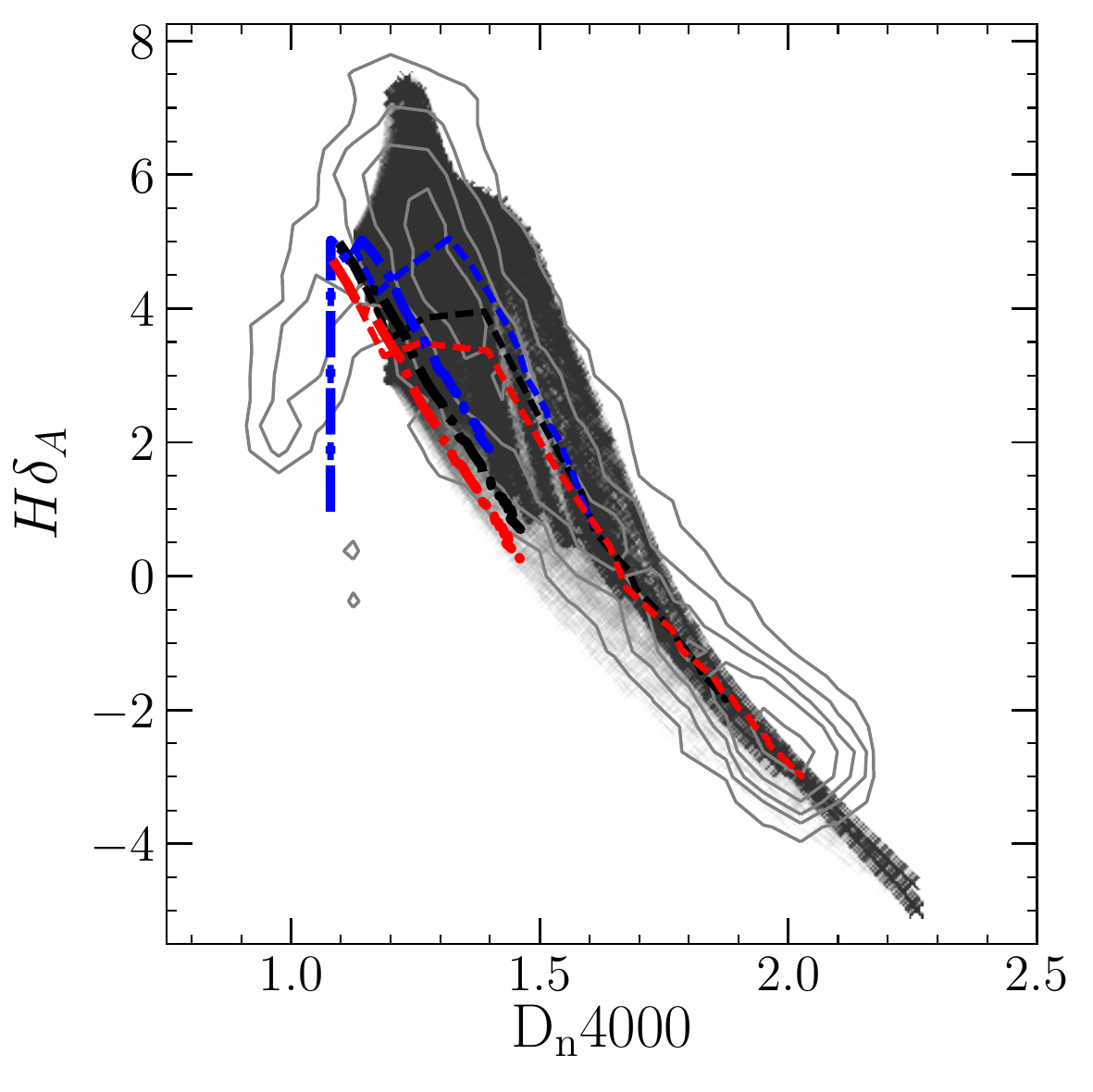}
\includegraphics[width=0.24\textwidth]{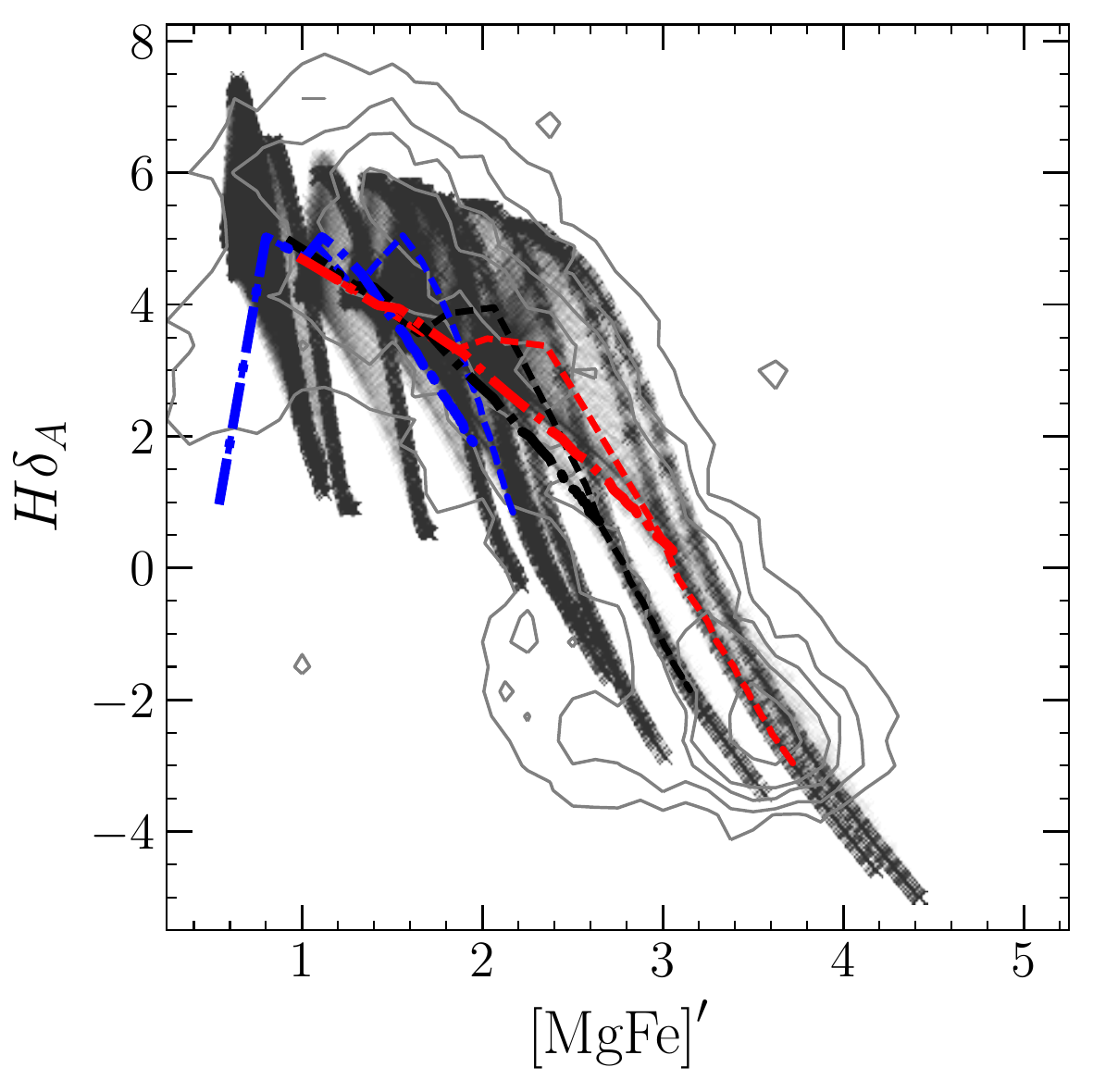}
\includegraphics[width=0.24\textwidth]{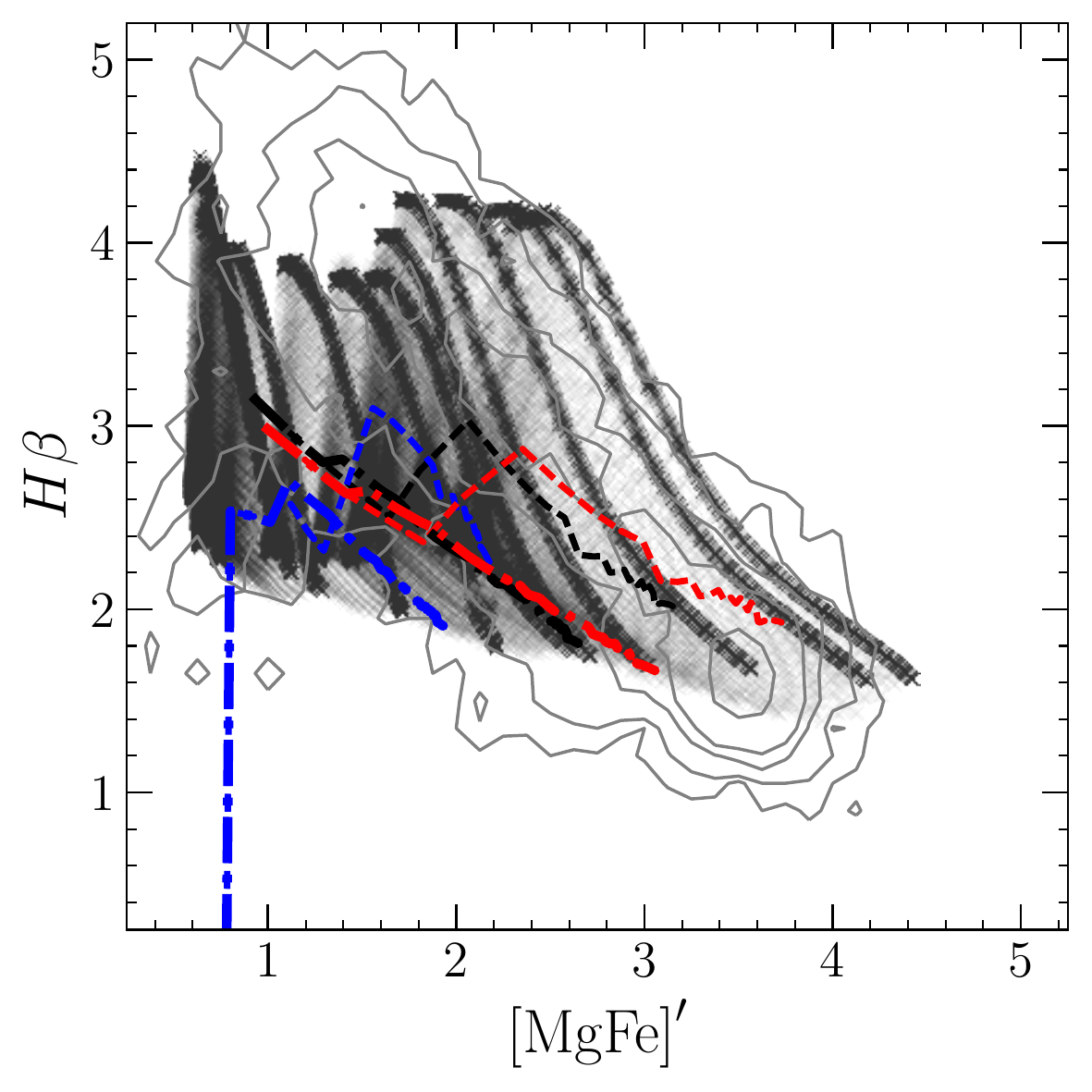}
\caption{Consistency test between actual spectral parameter measurements of the central spaxels (with $R/R_e < 0.1$) of all MPL-6 MaNGA galaxies (grey contours) and those measured from the synthetic spectra generated for the look-up table (transparent black crosses; see Section~\ref{sec:emcee}). The contours enclose $(11, 39, 68, 86, 96)\%$ of the spaxel measurements in each panel. We also show the tracks across cosmic time for a synthetic spectrum with constant SFR (thick dot-dashed lines) and for a synthetic spectrum with model quenching parameters $[\rm{t}_q, \tau] = [10.0, 0.5]~\rm{Gyr}$ (thin dashed lines; a relatively rapid quench) for $0.2~\rm{Z}_{\odot}$, $1.0~\rm{Z}_{\odot}$ and $1.6~\rm{Z}_{\odot}$ metallicities in blue, black and red respectively. We have not attempted to recreate the distributions across spectral parameter space seen for this sample of real galaxy spectra (see Section~\ref{sec:poptest} and Figure~\ref{fig:mangacompare} for such a test), we are merely showing the spectral parameters for the set of quenching SFHs we have generated across the 4-dimensional look-up table (in which $t_{obs} > 11.85~\rm{Gyr}$, i.e. $z \lesssim 0.15$, rather than covering all of cosmic time like the tracks shown by the dashed and dot-dashed lines), which we have shown in Figure~\ref{fig:rainbow} are degenerate. See Section~\ref{sec:consistency}.}
\label{fig:compare_manga_specmeas}
\end{figure*}

\subsection{Consistency of spectral parameter measurements}\label{sec:consistency}

Before testing the performance of the code, we tested the consistency of the measurements of the predicted spectral parameters generated in the look-up table (see Section~\ref{sec:emcee} and Figure~\ref{fig:rainbow}). To do this, we collated the spectral parameters for all the central spaxels (with $R/R_e < 0.1$ to give a reasonable sample size) of all MPL-6 MaNGA galaxies using the Marvin interface developed for MaNGA \citep{cherinka18}. These are shown by the black contours in each of the panels of Figure~\ref{fig:compare_manga_specmeas}. Overlaid are points showing the spectral measurements for the synthetic FSPS spectra from the look-up table. We can see that similar ranges are found for the modelled SFHs as for the spaxels of real MaNGA galaxies, suggesting that the models produced are appropriately generated and measured. Note that we have not attempted to recreate the distributions across spectral parameter space seen for this sample of real MaNGA galaxy spectra (see Section~\ref{sec:poptest} for such a test). We are merely showing the spectral parameters for the set of quenching SFHs we have generated across the 4-dimensional look-up table (in particular where $t_{obs} > 11.85~\rm{Gyr}$, i.e. $z \lesssim 0.15$ rather than covering all of cosmic time). Therefore we do not expect to cover the full range in spectral parameters seen for the real MaNGA galaxy spectra, since these will also include spectra that are starbursting, have increasing star formation rates or contain younger stellar populations. Whereas \textsc{snitch} is specifically designed to target the properties of quenching stellar populations.

\subsection{Testing precision}\label{sec:precisiontest}


In order to test that \textsc{snitch} can find the correct quenched SFH model for a given set of spectral features, 25 synthesised galaxy spectra were created with known SFH parameters (i.e. known randomised values of $\theta = [Z, t_q, \log \tau]$) from which synthetic spectra were generated and predicted spectral features were measured (see Section~\ref{sec:fsps}). These were input into \textsc{snitch}, assuming a $10\%$ error on each spectral parameter measurement, to test whether the known values of $\theta$ were reproduced, within error, for each of the 25 synthesised galaxies. 
In all cases the true values reside within the parameter space explored by the walkers left over after pruning, which trace the global minimum of the posterior probability. \textsc{snitch} therefore succeeds in locating the true parameter values within the degeneracies of the SFH model for known values. However, the spread in the walker positions generally gets broader as the inferred $\tau$ value gets larger (i.e. longer quench) and the inferred $t_q$ value gets smaller (i.e. earlier quench). This is a product of both the logarithmic spacing in the look-up table generated for use in \textsc{snitch} (see Section~\ref{sec:emcee}) and an observational effect, since spectral signatures of a longer, earlier quench will have been washed out over time. 

This test demonstrates how \textsc{snitch} is precise in recovering the parameters describing the true SFHs, however that precision varies across the parameter space. The median difference between known and inferred parameter values for 25 random SFHs is $[\Delta Z,~\Delta t_q,~\Delta \tau]~=~[0.1~\rm{Z}_{\odot},~0.3~\rm{Gyr},~0.2~\rm{Gyr}]$ and the maximum difference between the inferred and true values are $[\Delta Z,~\Delta t_q,~\Delta \tau]~=~[0.7~\rm{Z}_{\odot},~3.7~\rm{Gyr},~1.4~\rm{Gyr}]$.

\begin{table*}
\centering
\caption{The mean uncertainties ($\pm1±\sigma$) on the best fit and difference in known and best fit values ($\Delta [Z, t_q, \tau]$) for the 10 synthesised galaxy spectra returned when each spectral feature is omitted in turn. The accuracy in determining the metallicity, $Z$, parameter is most affected by the removal of $\rm{[MgFe]}^{\prime}$ and $\rm{D}_{\rm{n}}4000$. The accuracy in determining the time of quenching, $t_q$, parameter is most affected by the removal of $\rm{H}\beta$, $\rm{H}\delta_A$ and $\rm{D}_{\rm{n}}4000$. The accuracy in determining the rate of quenching, $\tau$, parameter is most affected by the removal of $\rm{H}\delta_A$, $\rm{EW}[\rm{H}\alpha]$ and $\rm{D}_{\rm{n}}4000$. See Section~\ref{sec:missingtest}.}
\label{table:missingtestone}
\begin{tabular*}{0.9\textwidth}{r@{\extracolsep{\fill}}|ccccccc}
Spectral feature omitted ~          & None & $\rm{H}\alpha$ & $\rm{D}_{\rm{n}}4000$ & $\rm{H}\beta$ & $\rm{H}\delta_A$ & $\rm{[MgFe]}^{\prime}$ \\ \hline
Average uncertainty, Z $1\sigma$~ &  $0.2$ &  $0.3$ &  $0.2$ & $0.2$ & $0.2$ & $0.4$ \\
Average uncertainty, $t_q$ $1\sigma~$ & $1.1$ & $1.9$ & $1.7$ & $2.1$ & $3.2$ & $2.4$ \\
Average uncertainty, $\tau$ $1\sigma~$ & $0.4$ & $0.8$ & $0.9$ & $0.5$ & $0.5$ & $0.8$ \\ \hline
$\Delta Z~\rm{[Gyr]}~$  &  $0.1$  & $0.1$ & $0.3$ & $0.2$ & $0.1$ & $0.3$ \\
$\Delta t_q~\rm{[Gyr]}~$ &  $0.3$  & $1.3$ & $1.6$ & $1.5$ & $1.9$ & $0.8$ \\
$\Delta \tau~\rm{[Gyr]}~$ & $0.2$ & $2.3$ & $1.4$ & $2.1$ & $2.6$ & $1.5$
\end{tabular*}
\end{table*}

\subsection{Testing precision when less spectral information provided}\label{sec:missingtest}

\textsc{snitch} is designed so that not all of the spectral features have to be provided for the code to return an inferred quenching history. This is a particularly useful feature if the user is unable to obtain or measure a certain spectral feature. For example, if measurements are being obtained from archival data or a feature lies outside of the wavelength range of their spectrum. 

Users should note that quenching histories inferred given fewer inputs results in a larger uncertainty on the quoted best fit parameters returned by \textsc{snitch}. To quantify this we generated 10 random $[Z, t_q, \log \tau]$ values and used them to generate synthetic spectra, in which the predicted spectral features were measured and used as inputs to \textsc{snitch}, each time omitting one of the spectral features from the list of inputs. The mean uncertainties on the best fit and difference between known and best fit values returned when each spectral feature is omitted are quoted in Table~\ref{table:missingtestone}. The accuracy in determining the metallicity, $Z$, parameter is most affected by the removal of $\rm{[MgFe]}^{\prime}$ and $\rm{D}_{\rm{n}}4000$. The accuracy in determining the time of quenching, $t_q$, parameter is most affected by the removal of $\rm{H}\beta$, $\rm{H}\delta_A$ and $\rm{D}_{\rm{n}}4000$. The accuracy in determining the rate of quenching, $\tau$, parameter is most affected by the removal of $\rm{H}\delta_A$, $\rm{EW[H}\alpha]$ and $\rm{D}_{\rm{n}}4000$.

For further combinations of missing parameters, we suggest the user completes their own tests to determine how the quoted uncertainty will change with the omission of more than one spectral feature. However, we do not recommend using \textsc{snitch} if the number of available spectral features is less than $4$. If this is the case, the number of inputs given to the code will be equal to or less than the number of parameters to be inferred and the resulting SFH will be unreliable. 

\subsection{Population testing}\label{sec:poptest}

\begin{figure*}
\centering
\includegraphics[width=\textwidth]{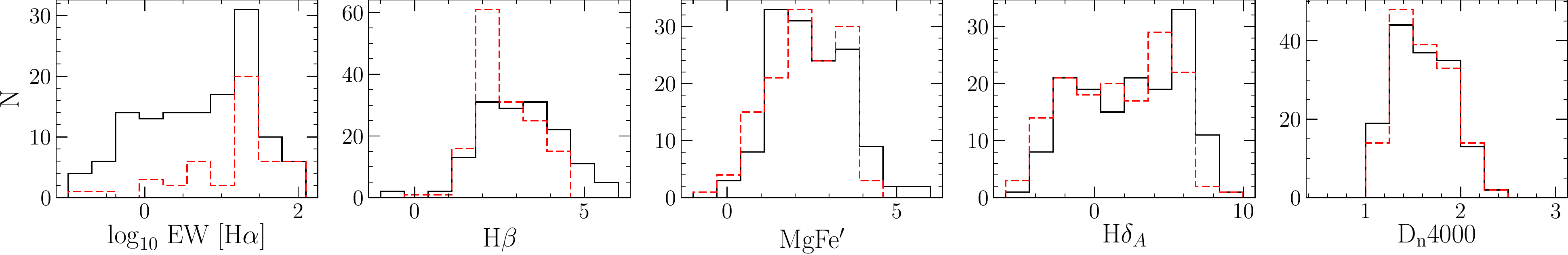}
\caption{The distribution, from right to left of the $\log_{10}$EW$[\rm{H}\alpha]$, $\rm{H}\beta$, $\rm{[MgFe]}^{\prime}$, $\rm{H}\delta_A$ and $\rm{D}_{\rm{n}}4000$ values of a random spaxel in each of $150$ randomly selected observed MaNGA galaxies (black solid line). In each panel the distribution of the \textsc{snitch} inferred spectral parameter is shown by the red dashed line. See Section~\ref{sec:accuracytest}}
\label{fig:mangacompare}
\end{figure*}

A further test of \textsc{snitch} is to determine whether the inferred SFH parameters, $[Z,~t_q,~\log\tau]$, can reproduce the distribution of observed spectral features of a sample of galaxy spectra. We randomly selected a spaxel from each of $150$ MaNGA MPL-6 galaxies  and used the observed spectral parameters as inputs to \textsc{snitch}. We then used the inferred SFH parameters returned by \textsc{snitch} to estimate inferred spectral parameters for each of the $150$ galaxy spaxels. In order to add noise to these inferred spectral parameters, we also added a random multiple of the error on the observed spectral features, drawn from a Gaussian distribution with mean of $0$ and standard deviation of $1.1$ (i.e. normally distributed between roughly $-3$ and $3$ so that the noise added to the inferred value is distributed between $\pm3\sigma$).

Figure~\ref{fig:mangacompare} shows the distributions of the inferred and measured spectral parameters and highlights how the inferred values trace the original measured values well, in particular for the absorption features. Once again demonstrating that \textsc{snitch} can return a precise SFH for a galaxy spectrum. 

However, we can see that \textsc{snitch} appears to struggle to reproduce the distribution of $\log_{10}$EW$[\rm{H}\alpha]$. This is due to the fact that the look-up tables which \textsc{snitch} uses are masked for $\log_{10}$EW$[\rm{H}\alpha]~\lesssim~1$ (see left most panel of Figure~\ref{fig:rainbow}) as these values become unreliable measurements due to the contamination from the nearby $[NII]$ doublet. Therefore the inferred SFH found by \textsc{snitch} will have a null EW$[\rm{H}\alpha]$ value where star formation is minimal. This was true for $92$ of the $150$ galaxy spaxels and so these values are not plotted in the distribution shown by the red curve in the left most panel of Figure~\ref{fig:mangacompare}. We did not mask the observed EW$[\rm{H}\alpha]$ values in order to provide \textsc{snitch} with five inputs for each MaNGA galaxy, as a control, so these values are still shown in the distribution shown by the black curve in the left most panel of Figure~\ref{fig:mangacompare}. 

\subsection{Testing accuracy}\label{sec:accuracytest}

We have shown in the previous section that \textsc{snitch} can return precise known values for SFHs, however now we must test its accuracy. In order to quantify this we have run \textsc{snitch} on spectra which have previously derived SFHs. Firstly, on those which have had similar simple models derived (Section~\ref{secsec:compare}) and then on spectra with SFHs from hydrodynamic simulations (Section~\ref{sec:lgalaxies}).

\begin{table*}
\centering
\caption{The mean star formation fraction (SFF) in each age bin for the six galaxy samples quoted by \protect\cite[][TSFF]{tojeiro13} and returned by \textsc{snitch}. Each value is quoted with an uncertainty, for the \protect\cite{tojeiro13} values this is quoted as the standard error on the mean for each bin with the same precision as \protect\citeauthor{tojeiro13} quote in their Table 2. For the \textsc{snitch} values the uncertainty stated is calculated from the SFH parameters at the $16$th and $84$th walker positions (see Section~\ref{sec:emcee}) and are quoted to a the nearest whole number since the \textsc{snitch} uncertainties are much broader than the ones calculated by \protect\citeauthor{tojeiro13} The SFF and $1\sigma$ errors are given in units of $10^{-3}$. See Section~\ref{secsec:compare}.}
\label{table:tojeirocompare}
\begin{tabular*}{\textwidth}{l|cc|cc|cc|cc}
                        Look-back time & \multicolumn{2}{c|}{$0.01 - 0.074~\rm{Gyr}$}     & \multicolumn{2}{c|}{$0.074 - 0.425~\rm{Gyr}$} & \multicolumn{2}{c|}{$0.425 - 2.44~\rm{Gyr}$} & \multicolumn{2}{c}{$2.44 - 13.7~\rm{Gyr}$} \\ \hline
                        & TSFF & \textsc{snitch} SFF & TSFF          & \textsc{snitch} SFF          & TSFF          & \textsc{snitch} SFF         & TSFF         & \textsc{snitch} SFF        \\ \hline
Red ellipticals         & $0.11\pm0.047$   &  $1\pm_{1}^{1}$ &  $0.32\pm0.052$  &    $1\pm_{1}^{1}$   &  $33\pm1$   &   $2\pm_{2}^{13}$   &  $966\pm2.89$  &   $996\pm_{6}^{1}$   \\
Red ET spirals  & $0.65\pm0.45$   &  $10\pm_{9}^{19}$ &  $2.4\pm0.023$   &  $22\pm_{21}^{44}$     &  $36\pm3.8$   &   $244\pm_{241}^{488}$  &  $960\pm8.4$  &  $997\pm_{276}^{1}$    \\
Red LT spirals   &  $1.9\pm1.18$   &  $61\pm_{59}^{121}$  &  $5.6\pm0.0097$   &   $113\pm_{111}^{225}$    &  $59\pm12$   &   $315\pm_{311}^{630}$   &  $933\pm18.7$  &   $997\pm_{501}^{1}$    \\ \hline
Blue ellipticals        &  $2.5\pm1.3$   &  $108\pm_{107}^{217}$   &  $11\pm0.3$    &   $186\pm_{184}^{372}$    &  $52\pm11$   &   $319\pm_{315}^{637}$   &  $934\pm17.2$  &     $997\pm_{638}^{1}$ \\
Blue ET spirals &  $4.9\pm1.1$   &  $80\pm_{79}^{46}$  &  $14\pm0.14$    &   $134\pm_{133}^{74}$    &  $42\pm5.2$   &   $211\pm_{209}^{86}$   &  $938\pm9.2$  &  $554\pm_{217}^{437}$     \\
Blue LT spirals  &  $6.1\pm1.4$   &  $67\pm_{66}^{58}$  &  $11\pm0.34$   &   $113\pm_{109}^{94}$    &  $43\pm12$   &    $187\pm_{184}^{113}$  &  $939\pm19.3$  & $615\pm_{279}^{372}$                              
\end{tabular*}
\end{table*}

\subsubsection{Comparing with other SFH inference codes}\label{secsec:compare}

\begin{figure}
\centering
\includegraphics[width=0.48\textwidth]{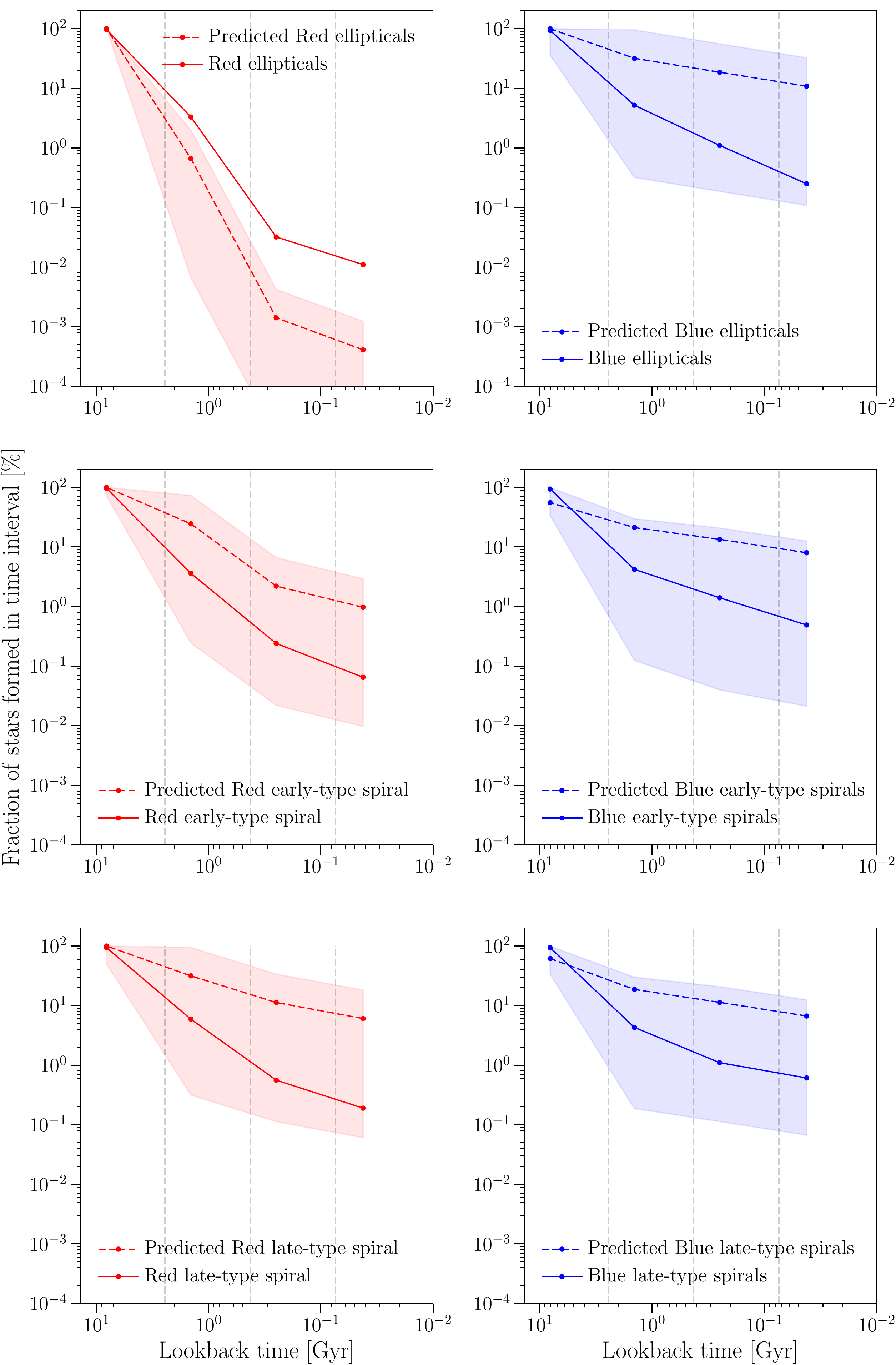}
\caption{The mean star formation fraction (SFF) in each age bin for the six galaxy samples analysed by \protect\cite[][solid lines]{tojeiro13} and returned by \textsc{snitch} (dashed lines). We have reproduced these plots in the exact same way as presented in Figure 7 of \citeauthor{tojeiro13} except that we have flipped the x-axis so that more recent epochs are on the right hand side for continuity with the rest of the figures in this work. The shaded region shows the $1\sigma$ error on the predicted SFF inferred by \textsc{snitch}. Note the logarithmic y-axis scale, given the large uncertainty in predicted SFFs inferred by \textsc{snitch}. These results suggest that \textsc{SNITCH} does return an accurate parametrised model of SFH, however when \textsc{snitch} is less accurate in its inference of the SFH this is reflected in the large uncertainties returned. See Section~\ref{secsec:compare}. }
\label{fig:tojeirocompare}
\end{figure}

In the case of the previously fitted simple SFH models, we have compared the results of \textsc{snitch} with the parametrised SFHs derived by \cite{tojeiro13} for $6$ stacked SDSS spectra of $13959$ red ellipticals, $381$ blue ellipticals, $5139$ blue late-type (LT) spirals, $294$ red LT spirals, $1144$ blue early-type (ET) spirals and $1265$ red ET spirals\footnote{Unfortunately \citeauthor{tojeiro13} did not select a separate sample of `green valley' galaxies, which have long been considered as the `crossroads' of galaxy evolution currently undergoing quenching between the blue cloud and red sequence \citep{smethurst15}. The `green' galaxies are therefore spread across the \protect\citeauthor{tojeiro13} red and blue samples. }. We measured the spectral features of each of the $6$ stacked spectra using the method outlined in Section~\ref{sec:dap} and input them into \textsc{snitch}. Since \citeauthor{tojeiro13} quoted their results in terms of the fraction of stars formed (SFF) in a given time period, we have followed the same method. In Table~\ref{table:tojeirocompare} we have listed the SFF for the six samples found by \citeauthor{tojeiro13} and the SFF for the best fit parameters inferred by \textsc{snitch} along with the uncertainty. These results are also plotted in Figure~\ref{fig:tojeirocompare}, recreating Figure 7 of \citeauthor{tojeiro13}

We can see from these results that \textsc{snitch} broadly agrees with the results of \citeauthor{tojeiro13}, within the uncertainties. However, the uncertainties returned by \textsc{snitch} are much broader for blue galaxies, particular for ET spirals, as seen in Figure~\ref{fig:tojeirocompare}. This is to be expected since \textsc{snitch} fits a quenching SFH model to a galaxy spectrum and so would return a less accurate SFH for star forming spectra (see Section~\ref{secsec:starforming}). There is also some discrepancy between the recent SFFs inferred by \textsc{snitch} and quoted by \citeauthor{tojeiro13} for the red ellipticals. This is presumably because of the incredibly small SFFs occurring at these recent epochs, which are difficult to constrain. Quenching must therefore have occurred at early epochs in these red ellipticals, which will dilute the spectral features giving rise to an uncertain fit.  These results suggest that \textsc{snitch} does return an accurate parametrised model of SFH at least for galaxies which are currently quenching or recently fully quenched (within at least the last $\sim2.5~\rm{Gyr}$, i.e. $z \lesssim 0.2$), however when \textsc{snitch} is less accurate in its inference of the SFH this is reflected in the large uncertainties returned.   

\begin{figure}
\centering
\includegraphics[height=0.91\textheight]{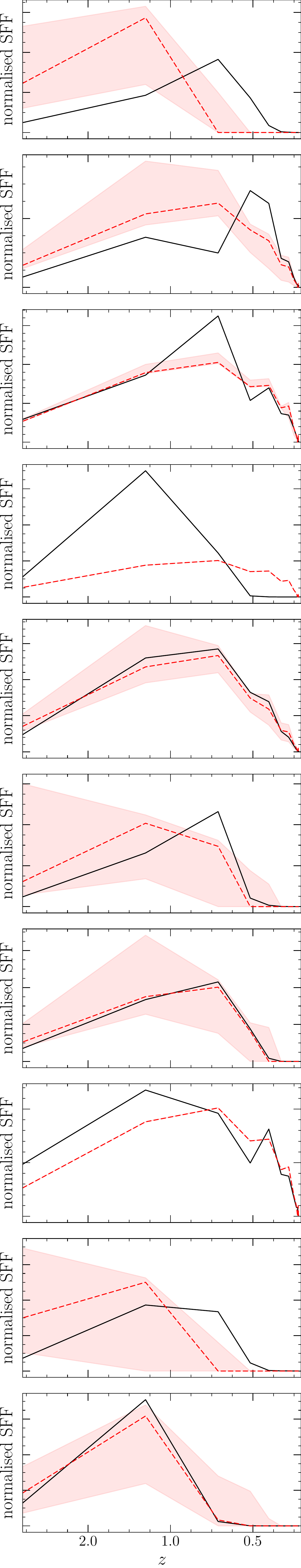}
\caption{Comparison of the SFFs generated by \texttt{Lgalaxies} (black) and those inferred by \textsc{snitch} (red, dashed with shaded uncertainty regions; note that two panels have very small uncertainties) for $10$ randomly selected synthetic spectra with SFRs in the range $0~<~\rm{SFR}~[M_{\odot}~\rm{yr}^{-1}]~<~1$, and stellar masses $10^9~<~M_{*}~[M_{\odot}]~<~10^{11}$. Note how \textsc{snitch} is sensitive to the most recent change in the SFF. See Section~\ref{sec:lgalaxies}.}
\label{fig:lgalsfhs}
\end{figure}

\begin{figure}
\centering
\includegraphics[width=0.49\textwidth]{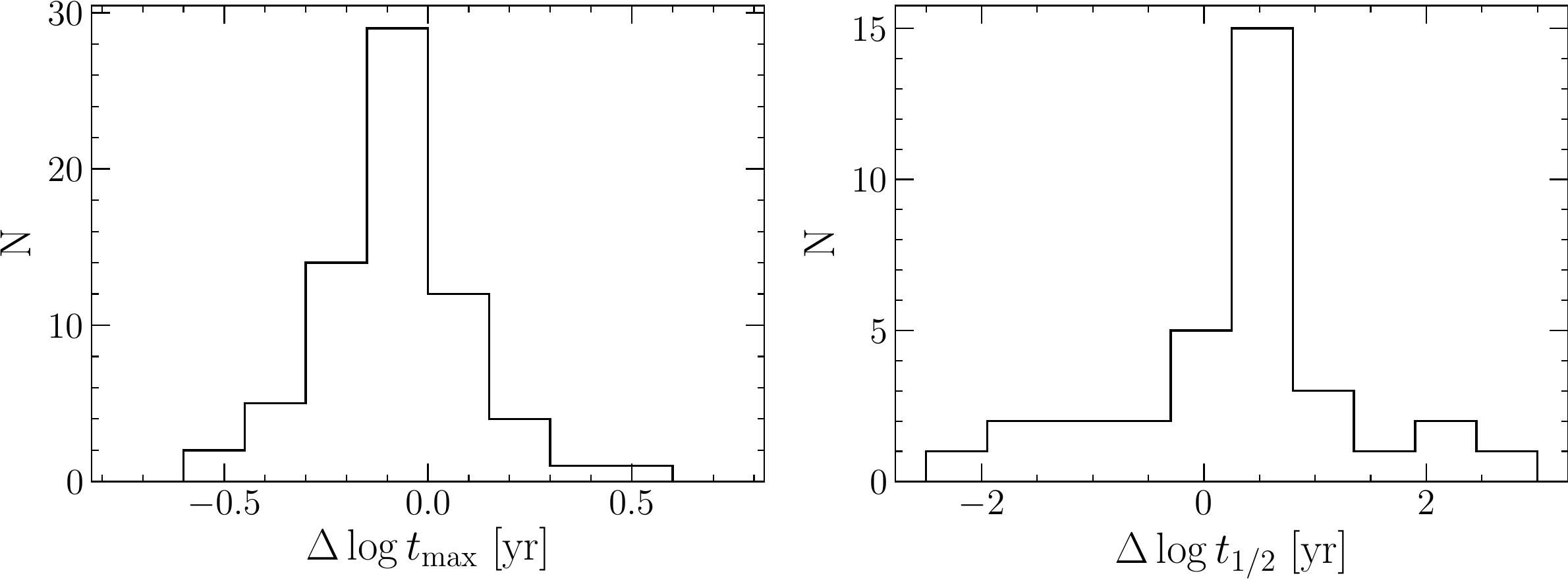}
\caption{Comparison of the difference between the calculated and inferred time of maximum SFR ($\Delta \log t_{\rm{max}}$; left) and time for the SFR to drop to half of the maximum value ($\Delta \log t_{1/2}$; right) for the $104$ synthetic SFHs with a non zero quasar accretion rate generated by \texttt{LGalaxies}. See Section~\ref{sec:lgalaxies}.}
\label{fig:genlgalsfhs}
\end{figure}



\subsubsection{Comparing with known SFHs from hydrodynamic simulations}\label{sec:lgalaxies}

We generated $8238$ simulated galaxy SFHs using the \texttt{LGalaxies} suite of hydrodynamic simulations \citep{henriques15}\footnote{These simulated SFHs were kindly generated by R. Asquith at the University of Nottingham.} at a redshift of $z=0.043$ (the mean redshift of the MaNGA DR14 sample) with a range of SFRs, $0 < \rm{SFR}~[M_{\odot}~\rm{yr}^{-1}] < 1$, and stellar masses $10^9 < M_{*} [M_{\odot}] < 10^{11}$. Of these $8238$ simulated galaxies we selected all of those flagged by \texttt{LGalaxies} to have a quasar accretion rate above zero\footnote{The development of this code has been driven by the desire to study the effects of AGN feedback on the SFHs of galaxies. This threshold on the quasar accretion rate was applied in order to supplement further study and comparison with observations in future work. It also doubled as a convenient way of limiting the sample size in this test of \textsc{snitch}.}. This resulted in $104$ simulated galaxy SFHs. We used the FSPS models of \cite{conroy10} to generate synthetic spectra for each of these $104$ simulated SFHs (as described in Section~\ref{sec:fsps}) and then measured their spectral features using the MaNGA DAP functions outlined in Section~\ref{sec:dap}. We then input these measurements into \textsc{snitch} to derive the best fit $[Z, t_q, \log \tau]$ parameters for our simple model of SFH to compare with the known SFH output by the hydrodynamic simulation. This test is therefore very similar to our tests with different known SFHs that we generated in Section~\ref{sec:precisiontest}, however the SFHs generated by the hydrodynamic simulation can be classed as both more varied and more characteristic of real galaxy SFHs in this case.

Figure~\ref{fig:lgalsfhs} shows the normalised SFFs as generated by \texttt{Lgalaxies} and inferred for their spectra by \textsc{snitch} for $10$ randomly selected simulated SFHs. We can see that the output from \textsc{snitch} largely agrees, within the uncertainties, with the known SFHs of Lgalaxies. Although not all details of the \texttt{Lgalaxies} SFHs are reproduced, \textsc{snitch} identifies the most recent epoch with a dramatic change in the SFR.

We can also generalise the SFHs generated by \texttt{Lgalaxies} and inferred by \textsc{snitch} into two parameters, the time of maximum SFR, $\log t_{\rm{max}}$, and the time for the SFR to drop to half of the maximum value ($\log t_{1/2}$). Note, that if a galaxy's SFR is increasing then we cannot derive a value for $t_{1/2}$. These generalised parameters roughly trace the exponential SFH parameters of $t_q$ and $\tau$, but allow for a comparison to the SFHs generated by \texttt{Lgalaxies} which are not constrained to an analytic form. Figure~\ref{fig:genlgalsfhs} shows the difference between the generated and inferred values of $\log t_{\rm{max}}$ \& $\log t_{1/2}$. We can see that for the majority of synthetic spectra the inferred SFH parameters are comparable to those generated by \texttt{Lgalaxies}. However there is a much larger spread in $\Delta \log t_{1/2}$ (shown in the right panel of Figure ~\ref{fig:genlgalsfhs}) than in $\Delta \log t_{\rm{max}}$ (shown in the left panel), suggesting that for galaxies with more complex SFHs, \textsc{snitch} will return a more accurate value for the time of quenching, $t_q$, than for the rate that quenching occurs, $\log \tau$.

\subsection{Testing performance with different SFH definitions}\label{sec:alldiffSFHs}

\subsubsection{Star Forming SFHs}\label{secsec:starforming}

\begin{figure}
\centering
\includegraphics[width=0.475\textwidth]{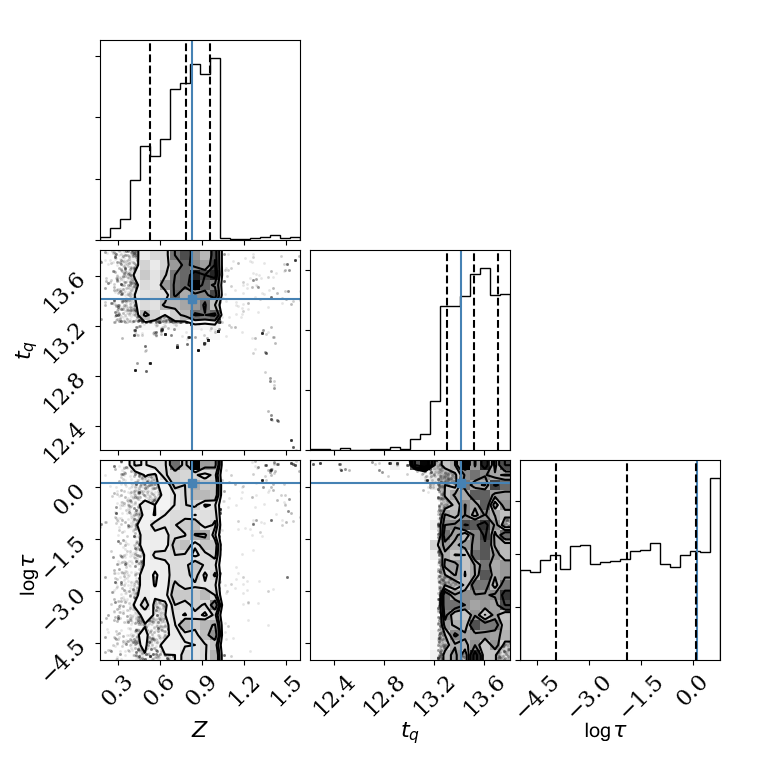}
\caption{Example output from \textsc{snitch} showing the posterior probability function traced by the MCMC walkers across the three dimensional parameter space $[Z,t_q,\log \tau]$, for a synthetic galaxy spectrum which is still star forming with a constant SFR.  Dashed lines show the 18th, 50th and 64th percentile of each distribution function which can be interpreted as the `best fit' with $1\sigma$. The blue lines show the known true values which \textsc{snitch} has managed to recover, within the uncertainties. Note that the walkers have explored only the parameter space where $t_q > t_{obs}$, i.e. the observed redshift of the galaxy (see Section~\ref{sec:fsps}), and all possible values of $\log \tau$, since the synthetic galaxy has not yet quenched and therefore all quenching rates are equally likely. See Section~\ref{secsec:starforming}.}
\label{fig:sfgal}
\end{figure}

We must also understand how \textsc{snitch} behaves when spectral parameters derived from a star forming galaxy spectrum are input. Figure~\ref{fig:sfgal} shows the example output from \textsc{snitch} across the three dimensional parameter space $[Z,t_q,\log \tau]$ for a synthetic galaxy spectrum which is still star forming at a constant rate at the time of observation, $t_{obs}$. Note that the walkers have explored only the parameter space where $t_q > t_{obs}$, i.e. the observed redshift of the galaxy (see Section~\ref{sec:fsps}), and all possible values of $\log \tau$, since the synthetetic galaxy has not yet quenched and therefore all quenching rates are equally likely.  

\subsubsection{Different Forms of Quenching SFHs}

Obviously, not all galaxies will be accurately described by an exponentially quenching SFH. In special use cases (for example studying post starburst galaxies) a different SFH may be defined by the user by replacing the \texttt{expsfh} function with their own. 

However, we have also tested how \textsc{snitch} behaves when spectra with known SFHs of different forms are input. We tested spectra with burst, many burst, normal and log-normal models of SFH, all of which are often used in the literature to model simple SFHs. 

We found that \textsc{snitch} was always sensitive to the most recent epoch of star formation or quenching. For the burst and many-burst models, \textsc{snitch} returns a constant SFR up until the peak of the last burst at which point quenching happens very rapidly. Similarly for the log normal and normal SFHs, \textsc{snitch} returns a best fit SFH with constant SFR until the peak of the normal at which point it declines at a rate comparable to the drop off of the Gaussian SFH. All of these tests suggest that \textsc{snitch} is most sensitive to the most recent epoch of star formation but can also roughly trace the quenching of star formation even if the true decline does not occur at an exponential rate. 


\section{Conclusions}

Given the recent influx of spectral data from integral field unit (IFU) surveys, there is need for a tool that allows a user to quickly derive a simple, informative star formation history (SFH) in order to compare the SFHs of spectra within a single IFU data cube or across a large population of galaxy spectra. We have therefore developed \textsc{snitch}, an open source \emph{Python} package which uses a set of five absorption and emission spectral features to infer the best fit parameters describing an exponentially declining model of SFH. To do this, \textsc{snitch} assumes a set of SFH parameters and convolves them with a stellar population synthesis (SPS) model to generate a synthetic spectrum. The predicted absorption and emission spectral features are then measured in this synthetic spectrum (using the same method developed to fit the observed spectra in MaNGA data cubes). The predicted spectral features for many different model SFHs are then compared to the input observed spectral features by \textsc{snitch} to find the best fit SFH model using Bayesian statistics and an MCMC method. \textsc{snitch} returns the best fit time of quenching, exponential rate of quenching and SPS model metallicity to the input spectral features. The typical run time for a single spectrum is around 2 minutes on a laptop machine. 

\textsc{snitch} was developed for specific use on the MaNGA IFU data cubes, however, it is fully customisable by the user for a specific science case, for example by changing the SFH model, spectral features used in the inference or the method used to measure spectral features in the synthetic spectra. We advocate for the use of \textsc{snitch} as a comparative tool within an IFU data cube or across a large population of spectra, rather than to derive a detailed SFH of a single spectra due to the generalising nature of the analytic SFH model.

We have demonstrated with rigorous testing that \textsc{snitch} is both precise and accurate at inferring the parameters describing an exponentially declining model of SFH. These tests suggest that \textsc{snitch} is sensitive to the most recent epoch of star formation but can also trace the quenching of star formation even if the true decline does not occur at an exponential rate.

\section*{Acknowledgements}

{\refereeii We thank the referee, I. Chilingarian, for insights which greatly helped to clarify the text.} The authors would like to thank A. Aragon-Salamanca for in depth discussions on the nature of galaxy spectral features and fitting methods. We would also like to thank R. Asquith for generating the \texttt{LGalaxies} simulation SFHs used in Section~\ref{sec:lgalaxies}.

RJS gratefully acknowledges research funding from the Ogden Trust and Christ Church, Oxford. 

This research made use of Marvin, a core Python package and web framework for MaNGA data, developed by Brian Cherinka, Jos\'e S\'anchez-Gallego, Brett Andrews, and Joel Brownstein. (MaNGA Collaboration, 2018).

Funding for the Sloan Digital Sky Survey IV has been provided by the Alfred P. Sloan Foundation, the U.S. Department of Energy Office of Science, and the Participating Institutions. SDSS acknowledges support and resources from the Center for High-Performance Computing at the University of Utah. The SDSS web site is \url{www.sdss.org}.

SDSS is managed by the Astrophysical Research Consortium for the Participating Institutions of the SDSS Collaboration including the Brazilian Participation Group, the Carnegie Institution for Science, Carnegie Mellon University, the Chilean Participation Group, the French Participation Group, Harvard-Smithsonian Center for Astrophysics, Instituto de Astrof\'isica de Canarias, The Johns Hopkins University, Kavli Institute for the Physics and Mathematics of the Universe (IPMU) / University of Tokyo, Lawrence Berkeley National Laboratory, Leibniz Institut für Astrophysik Potsdam (AIP), Max-Planck-Institut f\"ur Astronomie (MPIA Heidelberg), Max-Planck-Institut für Astrophysik (MPA Garching), Max-Planck-Institut f\"ur Extraterrestrische Physik (MPE), National Astronomical Observatories of China, New Mexico State University, New York University, University of Notre Dame, Observat\'orio Nacional / MCTI, The Ohio State University, Pennsylvania State University, Shanghai Astronomical Observatory, United Kingdom Participation Group, Universidad Nacional Aut\'onoma de M\'exico, University of Arizona, University of Colorado Boulder, University of Oxford, University of Portsmouth, University of Utah, University of Virginia, University of Washington, University of Wisconsin, Vanderbilt University and Yale University.

\bibliographystyle{mn2e}
\bibliography{refs}  

\end{document}